%% file: main.tex
\documentclass[tighten, times,twocolumn]{aastex631}

\usepackage{epsfig}
\usepackage{amsmath}
\usepackage{physics}
\usepackage{amssymb}
\usepackage{mathtools}
\usepackage{color}
\usepackage{natbib}
\usepackage{enumitem}
\usepackage{array}
\usepackage{makecell}

\newcommand{\bvec}[1]{{\mbox{{\boldmath$#1$}}}}	
\newcommand{\unitv}[1]{\bvec{\hat{#1}}}			

\input Figures.tex

\begin{document}

\title{Inertial Waves in a Nonlinear Simulation of the Sun's Convection Zone and Radiative Interior}

\author[0000-0001-9004-5963]{Catherine C. Blume}
\affiliation{JILA \& Department of Astrophysical and Planetary Sciences,
University of Colorado Boulder,
Boulder, CO 80309-0440, USA}

\author[0000-0001-7612-6628]{Bradley Hindman}
\affiliation{JILA \& Department of Astrophysical and Planetary Sciences,
University of Colorado Boulder,
Boulder, CO 80309-0440, USA}

\affiliation{Department of Applied Mathematics,
University of Colorado Boulder,
Boulder, CO 80309-0526, USA}

\author[0000-0001-9001-6118]{Loren I. Matilsky}
\affiliation{Department of Applied Mathematics,
University of California Santa Cruz,
Santa Cruz, CA 95064, USA}

\begin{abstract}
Recent observations of Rossby waves and other more exotic forms of inertial oscillations in the Sun’s convection zone have kindled the hope that such waves might be used as a seismic probe of the Sun’s interior. Here we present a 3D numerical simulation in spherical geometry that models the Sun’s convection zone and upper radiative interior. This model features a wide variety of inertial oscillations, including both sectoral and tesseral equatorial Rossby waves, retrograde mixed inertial modes, prograde thermal Rossby waves, the recently observed high-frequency retrograde (HFR) vorticity modes, and what may be latitudinal overtones of these HFR modes. With this model, we demonstrate that sectoral and tesseral Rossby waves are ubiquitous within the radiative interior as well as within the convection zone. We suggest that there are two different Rossby-wave families in this simulation that live in different wave cavities: one in the radiative interior and one in the convection zone. Finally, we suggest that many of the retrograde inertial waves that appear in the convection zone, including the HFR modes, are in fact all related, being latitudinal overtones that are mixed modes with the prograde thermal Rossby waves.
\end{abstract}


\input S1-Introduction.tex
\input S2-Simulation.tex
\input S3-SpectraSummary.tex

\input S4-RIWaves.tex

\input S5-CZWaves.tex
\input S6-Discussion.tex
\input S7-Conclusion.tex

\vspace{0.3cm}
We thank Keith Julien for clarifying conversations about mode nomenclature and Nick Featherstone for his help getting the simulation running.  C. Blume was primarily supported during this work by the University of Colorado Boulder George Ellery Hale Graduate Fellowship, along with the Future Investigators in NASA Earth and Space Sciences Technology (FINESST) award 80NSSC23K1624. L. Matilsky was supported by NSF award AST-2202253. This work was also supported by NASA Heliophysics through grants 80NSSC18K1125, 80NSSC19K0267, 80NSSC20K0193, and 80NSSC24K0271. Computational resources were provided by the NASA High-End Computing (HEC) Program through the NASA Advanced Supercomputing (NAS) Division at Ames Research Center. Rayleigh is hosted and receives support from the Computational Infrastructure for Geodynamics (CIG), which is supported by the National Science Foundation awards NSF-0949446, NSF-1550901, and NSF-2149126. This work was done in collaboration with the COFFIES DSC (NASA grant 80NSSC22M0162).

\input Z1-Linefitting.tex

\bibliography{Bibliography}

\end{document}

%% file: Figures.tex
\def\dr{
\begin{figure*}
{\includegraphics[width = 7in, keepaspectratio=true]{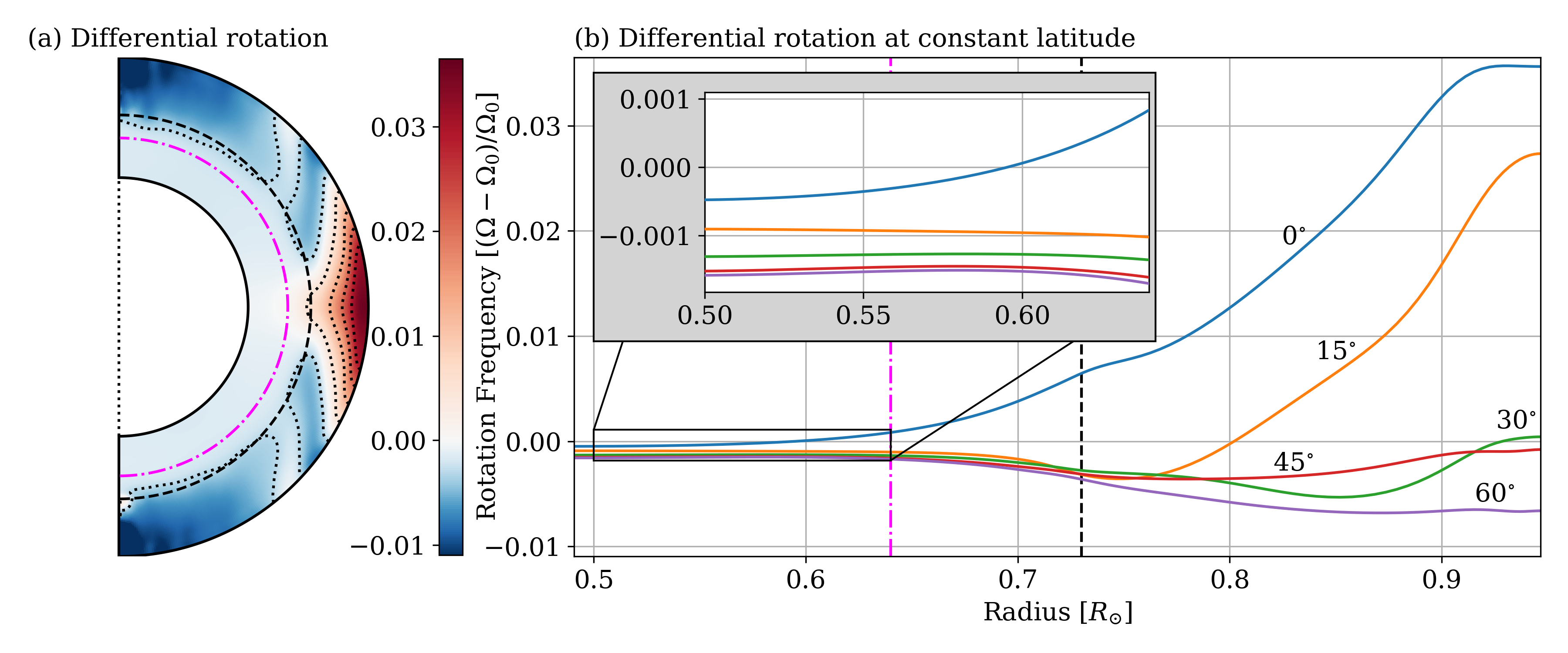}}

\vspace{-0.1in}
\caption{\normalsize  \textbf{Rotation profile achieved in our simulation}---This model features a differentially rotating convection zone and rigidly rotating radiative interior. (a) The azimuthally and temporally averaged rotation profile measured relative to the frame rate shown in a meridional cross-section. The north pole of the coordinate system is vertical on the page. The dashed black semi-circle indicates the base of the convection zone and the dash-dotted magenta semi-circle shows the bottom of the tachocline. (b) Radial cuts at several latitudes (indicated by labels) through the azimuthally averaged profile. The inset shows a magnified view of the rotation rate in the radiative interior, which has about $0.0015 \, \Omega_0$ of latitudinal differential rotation.  
\label{fig:diff_profile}
}
\end{figure*}
}

\def\simparams{
\begin{figure*}
{\includegraphics[width = 7in, keepaspectratio=true]{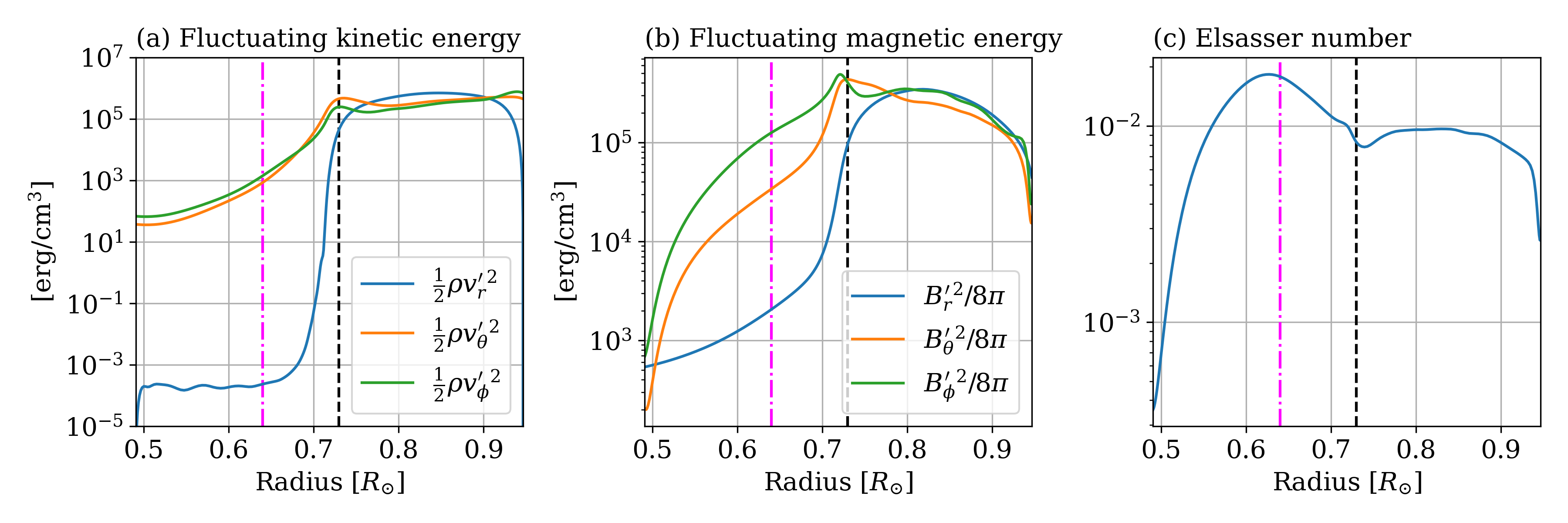}}

\vspace{-0.1in}
\caption{\normalsize  \textbf{Energy densities achieved in the simulation}---The velocity and magnetic field are primarily horizontal within the radiative interior. (a) Spherically-averaged fluctuating kinetic energy density separated into the contributions from each velocity component. Fluctuating velocities are defined by $v_i^\prime \equiv \langle (v_i - \langle v_i \rangle)^2 \rangle^{1/2}$, where $\langle\rangle$ denotes a temporal and spherical average over the equilibrated state. The horizontal contributions are orders of magnitude larger than the radial contribution within the radiative interior.  (b) Spherically-averaged magnetic energy density, where the fluctuating magnetic field components are defined in the same rms manner as the velocity fluctuations. The horizontal contributions of the magnetic field again dominate the vertical contribution, though this dominance is less than for the velocity field. (c) Dynamic Elsasser number, as defined by Equation~(\ref{eqn:Elsasser}). The Elsasser number is everywhere tiny, so the inertial waves are largely a hydrodynamic phenomenon with magnetism only acting as a small perturbation. \label{fig:simparams}
}
\end{figure*}
}

\def\spectrasummary{
\begin{figure*}
{\includegraphics[width = 7in, keepaspectratio=true]{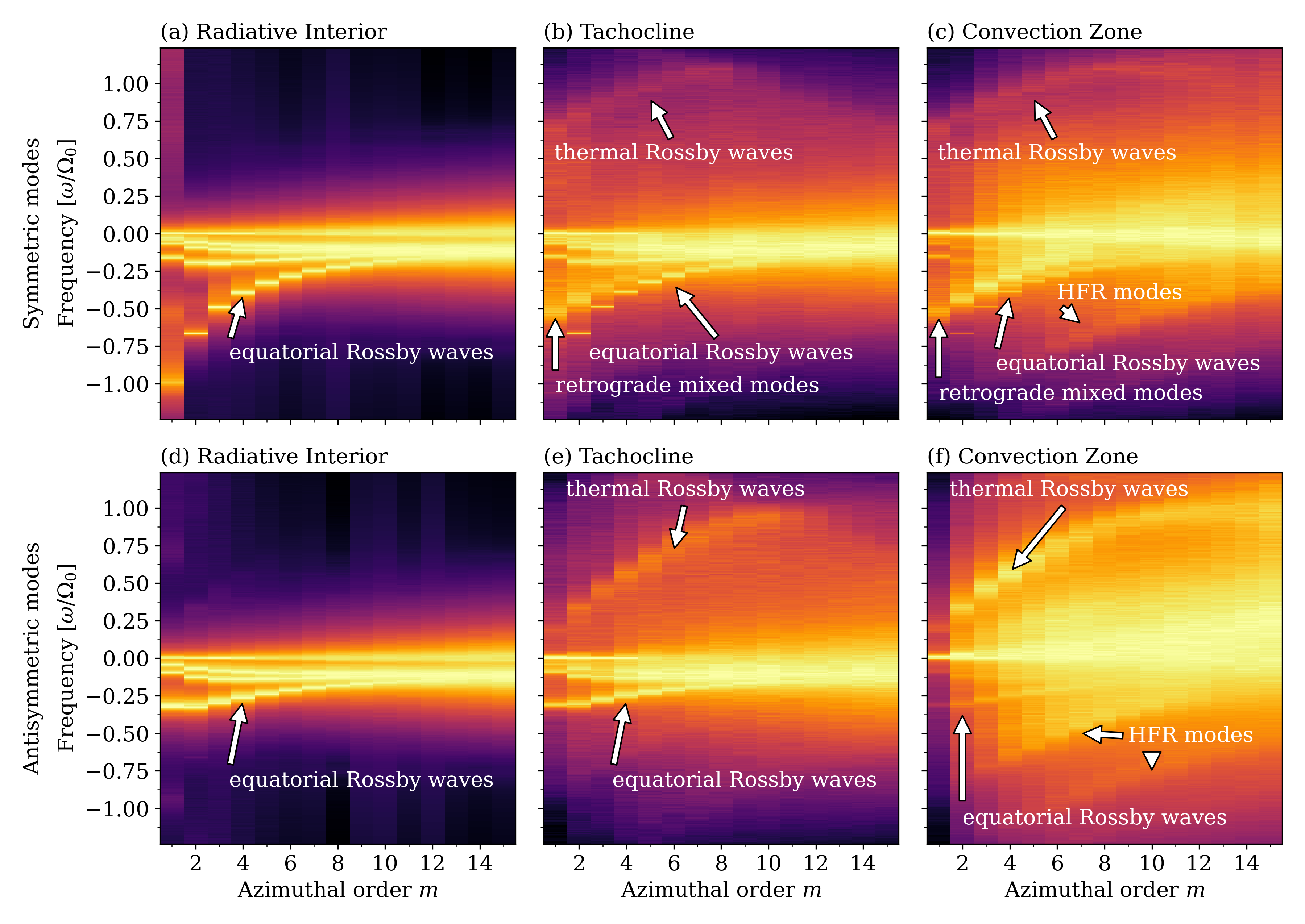}}

\vspace{-0.1in}
\caption{\normalsize  \textbf{Waves in the radiative interior, tachocline, and convection zone---}Power spectra of the radial vorticity as a function of azimuthal order $m$ and dimensionless temporal frequency $\omega/\Omega_0$, averaged over the radiative interior (left), tachocline (center), and convection zone (right). The top row displays the sum of power from modes that are symmetric about the equator with $\lambda = l - m = 0, 2, 4,$ and $6$, and the bottom row displays the sum of power from anti-symmetric modes with $\lambda = l - m = 1, 3,$ and $5$. The power is scaled separately for each $m$, so that the maximum power at any given $m$ has a value of unity. The images are shown with a logarithmic color table. We note a rich wavefield in both the radiative interior and convection zone. In particular, sectoral and tesseral Rossby waves are present throughout the domain, along with ridges of high-frequency power in the convection zone corresponding to HFR modes.
\label{fig:Wave spectra summary}
}
\end{figure*}
}

\def\rwradi{
\begin{figure}
{\includegraphics[width = 3in, keepaspectratio=true]{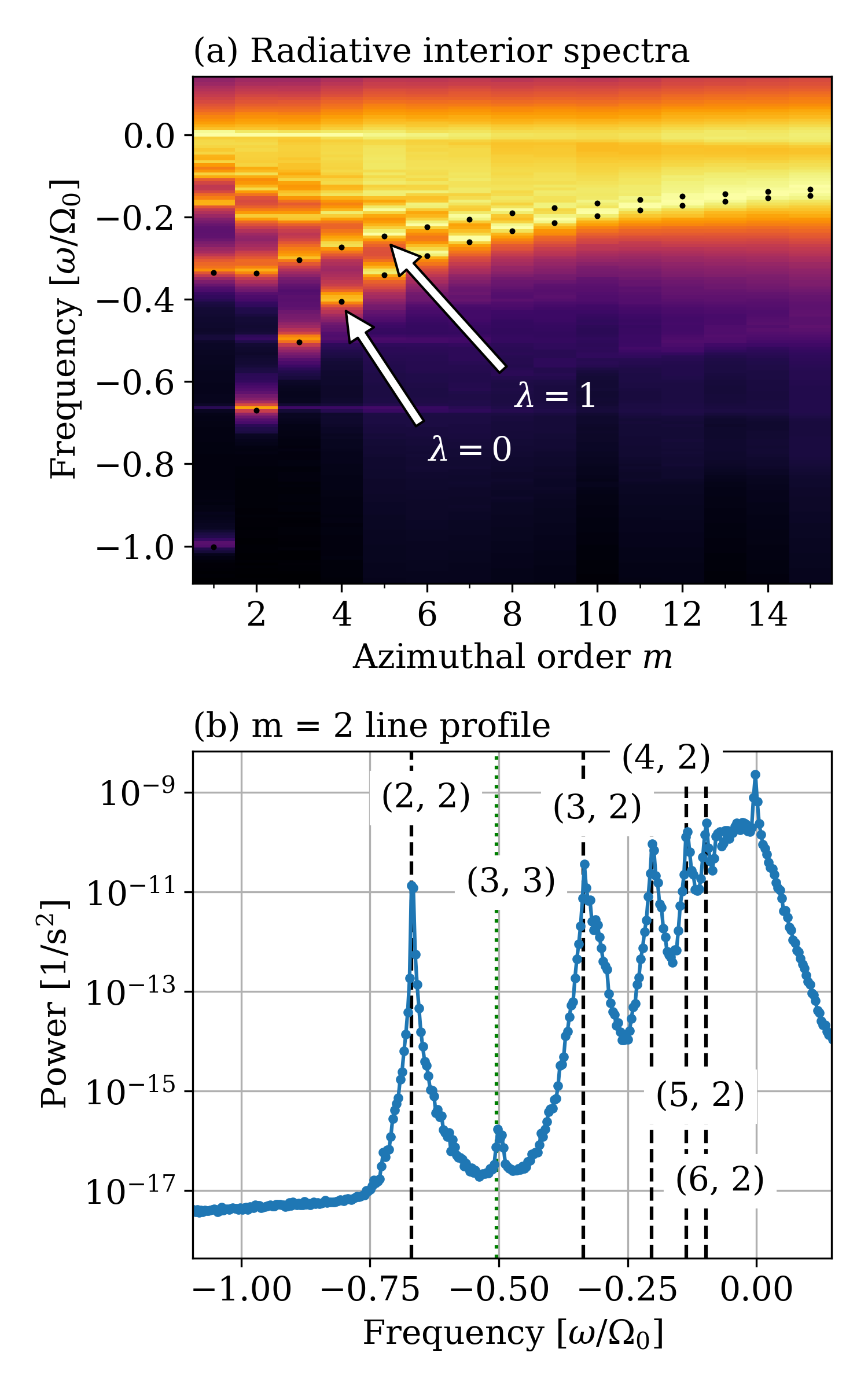}}

\vspace{-0.1in}
\caption{\normalsize \textbf{Equatorial Rossby waves in the radiative interior---}(a) Power spectra in radial vorticity as a function of azimuthal order $m$ and dimensionless temporal frequency $\omega/\Omega_0$, averaged over the radiative interior and summed over all latitudinal orders. The $\lambda = 0$ and $\lambda = 1$ ridges are labeled. The power is scaled separately for each $m$, so that the maximum power at any given $m$ has a value of unity, and shown with a logarithmic color table. The black dots indicate the theoretical values given by Equation ~(\ref{eq:disp}). A large number of both sectoral and tesseral equatorial Rossby modes are present throughout the radiative interior. (b) $m = 2$ power profile at radius $0.595 \, R_{\odot}$, summed over all $\lambda$, binned down in frequency by a factor of 32. The dashed black lines mark the theoretical values given by Equation ~(\ref{eq:disp}) for the labeled pairs $(l, m)$. The green dotted line denotes power contributed by a mode with a different value of $m$ (arising from a weak misalignment of the mean angular momentum vector and the spherical coordinate axis caused by numerical noise.) The modes blur together near zero frequency since their separation falls below the mode linewidths. Given that these peaks rise orders of magnitude above the noise floor, the vast majority of power in the radiative interior is concentrated in equatorial Rossby waves.
\label{fig:radispectra}
}
\end{figure}
}

\def\linecuts{
\begin{figure}
{\includegraphics[width = 3.25in, keepaspectratio=true]{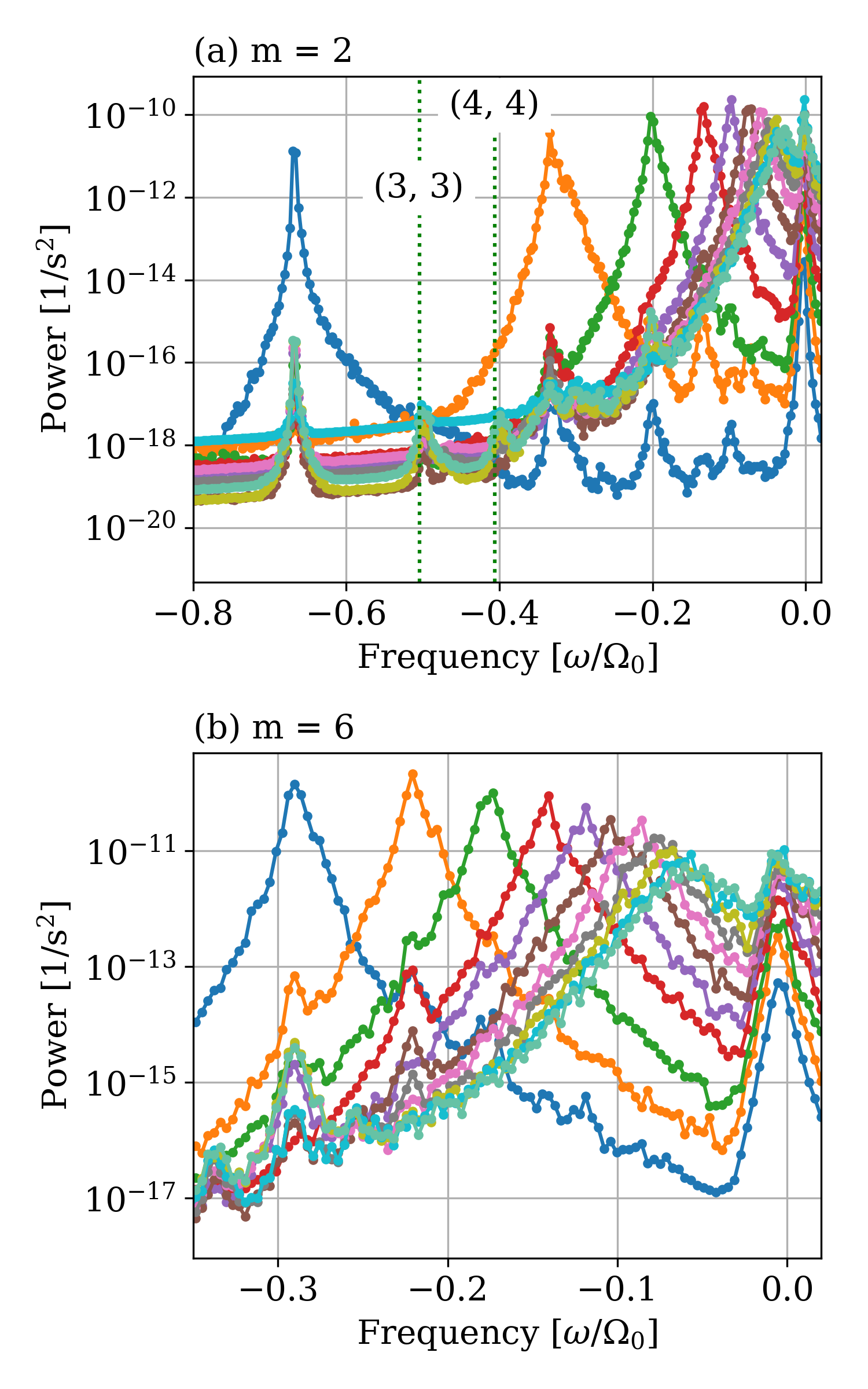}}

\vspace{-0.1in}
\caption{\normalsize  \textbf{Rossby-wave power profiles---}Power spectra of the radial vorticity as a function of dimensionless temporal frequency $\omega/\Omega_0$ for modes with $0 \leq \lambda \leq 10$ and (a) $m = 2$ and (b) $m = 6$. The power is taken at a radius of $0.595 \, R_{\odot}$ and binned down in frequency by a factor of 32. The peaks from left to right correspond to increasing values of $\lambda$ (or equivalently $l$). The higher $\lambda$ modes are clearly present, but when their power is summed together (as in Figure \ref{fig:radispectra}b), their frequency separation falls below their linewidths, and they become an indistinguishable blur. The green dotted lines denote power contributed by modes with a different value of $m$, labeled by ($l$, $m$).
\label{fig:linecuts}
}
\end{figure}
}

\def\eigenfunctions{
\begin{figure}
{\includegraphics[width = 3.25in, keepaspectratio=true]{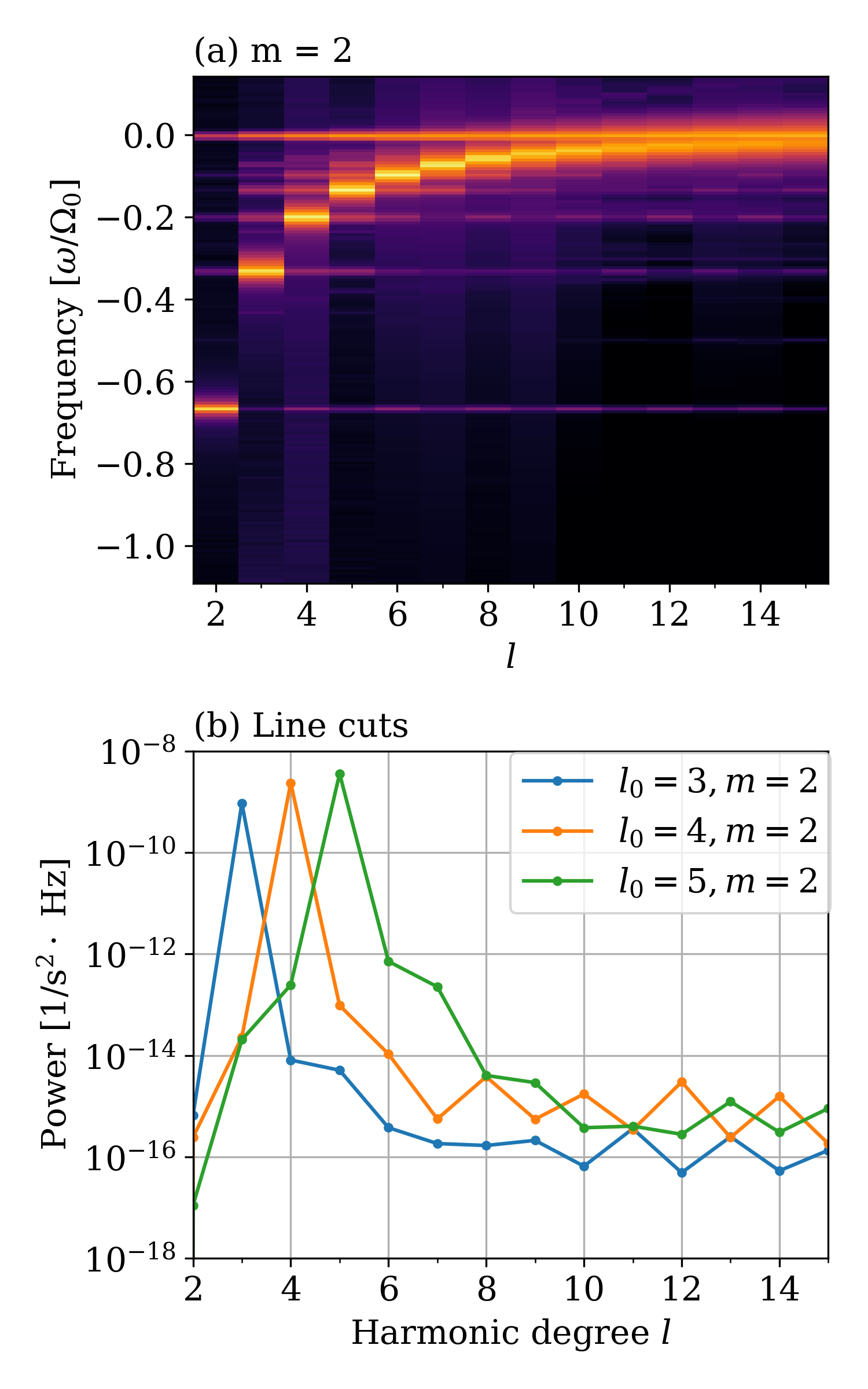}}

\vspace{-0.1in}
\caption{\normalsize  \textbf{Eigenfunction perturbations from asphericity---}Power for individual equatorial Rossby wave modes as a function of harmonic degree, demonstrating that the mode eigenfunctions are not a single spherical harmonic. (a) Radial vorticity power spectra for $m  = 2$ in the radiative interior. (b) Power as a function of harmonic degree $l$, summed over a narrow frequency window. The central frequency of the band is found by computing the mean of the power distribution, and the width of the window is $10 \sigma$, where $\sigma$ is the standard deviation. The power is clearly concentrated in a single spherical harmonic component with weak spread across nearby spherical harmonic components lying within $\pm 2$ of the principal component.
\label{fig:eigenfunctions}
}
\end{figure}
}

\def\czwaves{
\begin{figure*}
{\includegraphics[width = 7in, keepaspectratio=true]{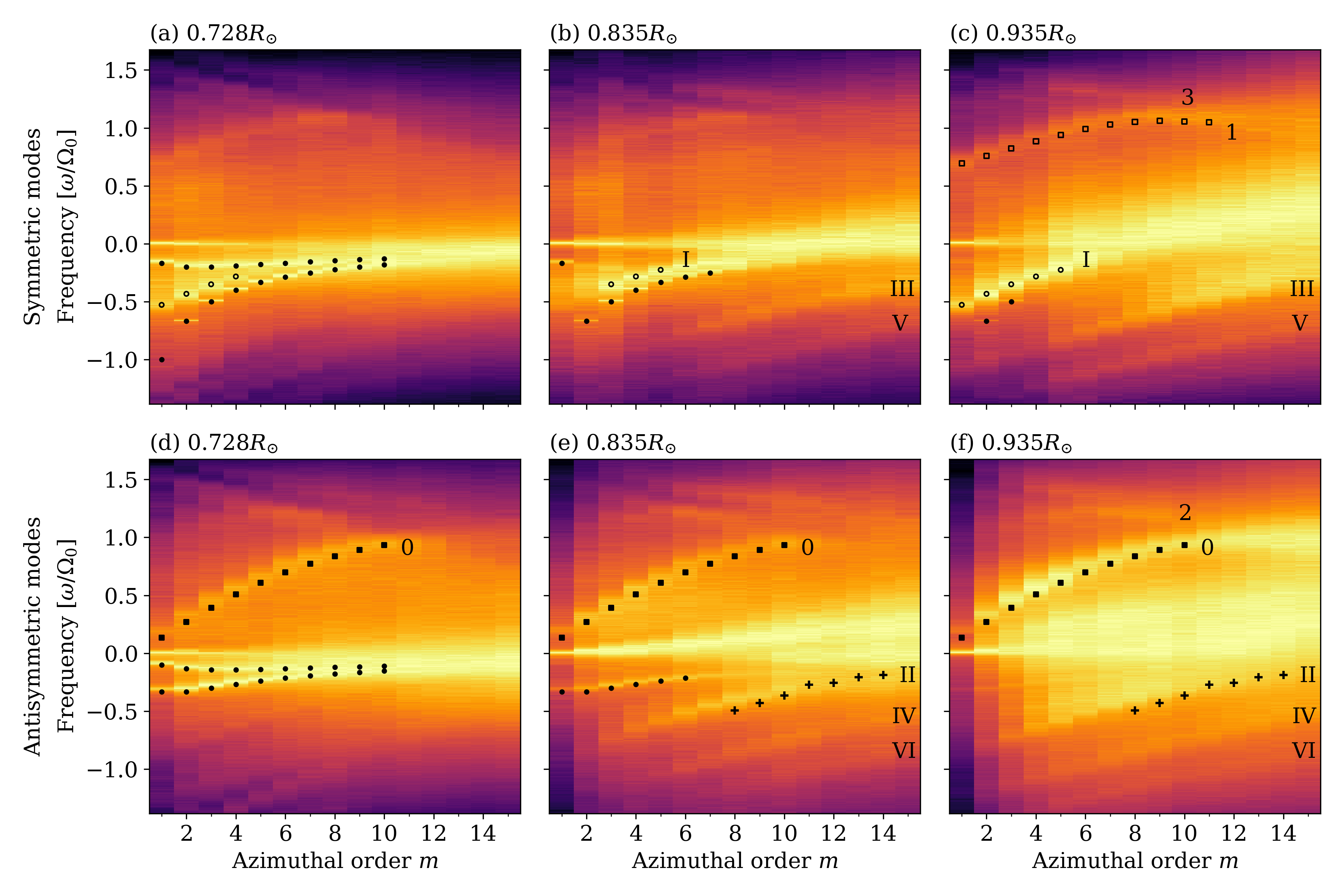}}

\vspace{-0.1in}
\caption{\normalsize  \textbf{Waves in the convection zone---}Power in radial vorticity as a function of azimuthal order $m$ and dimensionless temporal frequency $\omega/\Omega_0$ at three different radii within the convection zone. The top row displays the sum of symmetric modes $\lambda = 0$ and $2$, and the bottom row displays the sum of anti-symmetric modes $\lambda = 1$ and  $3$. The power is scaled separately for each $m$ so that the maximum power at any given $m$ has a value of unity. The images are shown on a logarithmic color table. The filled black circles are the expected values for equatorial Rossby waves from Equation~(\ref{eq:disp}) with no Doppler shift applied. Prograde thermal Rossby-wave ridges are numbered via their latitudinal node number in $v_{\phi}$, i.e., ``0" indicates the Busse modes (no latitudinal nodes), ``1" indicates the Roberts modes (one node), ``2" indicates modes with three latitudinally stacked convective columns (two nodes), and so on. Roman numerals label retrograde inertial waves, with I denoting the retrograde mixed mode and II-VI denoting HFR modes. The empty black circles and the empty black squares show the numerical frequencies from \citet{2022AA...662A..16B}. Filled black squares denote the semi-analytic dispersion relation for thermal Rossby waves from \citet{2022ApJ...932...68H}. The plus signs denote the observed HFR values from \citet{2022NatAs...6..708H}. A Doppler shift of $0.015 \, \Omega_0$ was applied to these four sets of values, as described in the text. This region features a wide variety of different inertial oscillations, some of which have been observed on the Sun and some of which have not (for a summary, see Table~\ref{tab:summarytable}).
\label{fig:czwaves}
}
\end{figure*}
}

\def\thermallines{
\begin{figure*}
{\includegraphics[width = 7in, keepaspectratio=true]{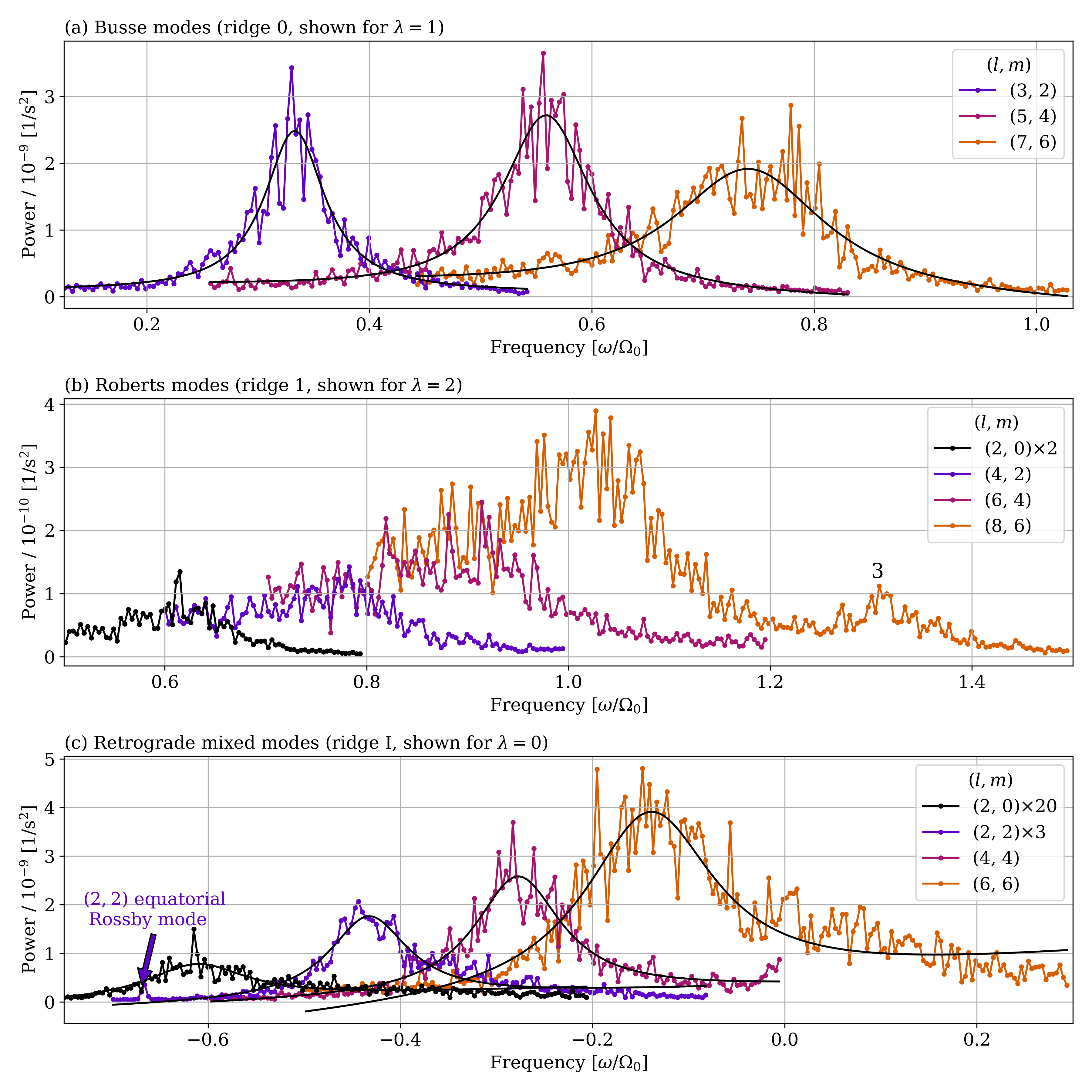}}

\vspace{-0.1in}
\caption{\normalsize  \textbf{Power profiles for thermal Rossby waves and mixed modes---}Selected power profiles from radial-vorticity power spectra, binned down in frequency by a factor of 32. (a) Busse modes, shown for $\lambda = 1$ at a radius of $0.889\,R_{\odot}$. (b) Roberts modes, shown for $\lambda = 2$, at a radius of $0.935 \, R_{\odot}$. The power in the $m = 0$ mode is multiplied by a factor of two for clarity. The bump labeled ``3" may indicate the presence of a thermal Rossby wave with three nodes in latitude in $v_{\phi}$. (c) The retrograde branch of the mixed mode of \citet{2022AA...662A..16B} shown for $\lambda = 0$ for $m = 2, 4,$ and $6$ and in $\lambda = 2$ for $m = 0$, at a radius of $0.935 \, R_{\odot}$. The power in the $m = 2$ mode is multiplied by a constant factor of 3 for clarity. The $(l, m) = (2, 0)$ mode, multiplied by a constant factor of 20, is plotted to show its relation to the $m = 0$ mode in (b). Lorentzian fits have been overplotted in black lines in Panels (a) and (c), and the corresponding values are listed in Tables \ref{tab:fitting} and \ref{tab:mixedfitting}, respectively. The thermal Rossby waves in (a) appear to be stable with a $Q$ factor of about 10. The prograde mixed modes in panel (b) and the retrograde mixed modes in panel (c) both have the $m=0$ mode illustrated, and it is clear that they have identical power profiles with $\omega\to-\omega$.
\label{fig:thermallinecuts}
}
\end{figure*}
}

\def\hfrlinecuts{
\begin{figure*}
{\includegraphics[width = 7in, keepaspectratio=true]{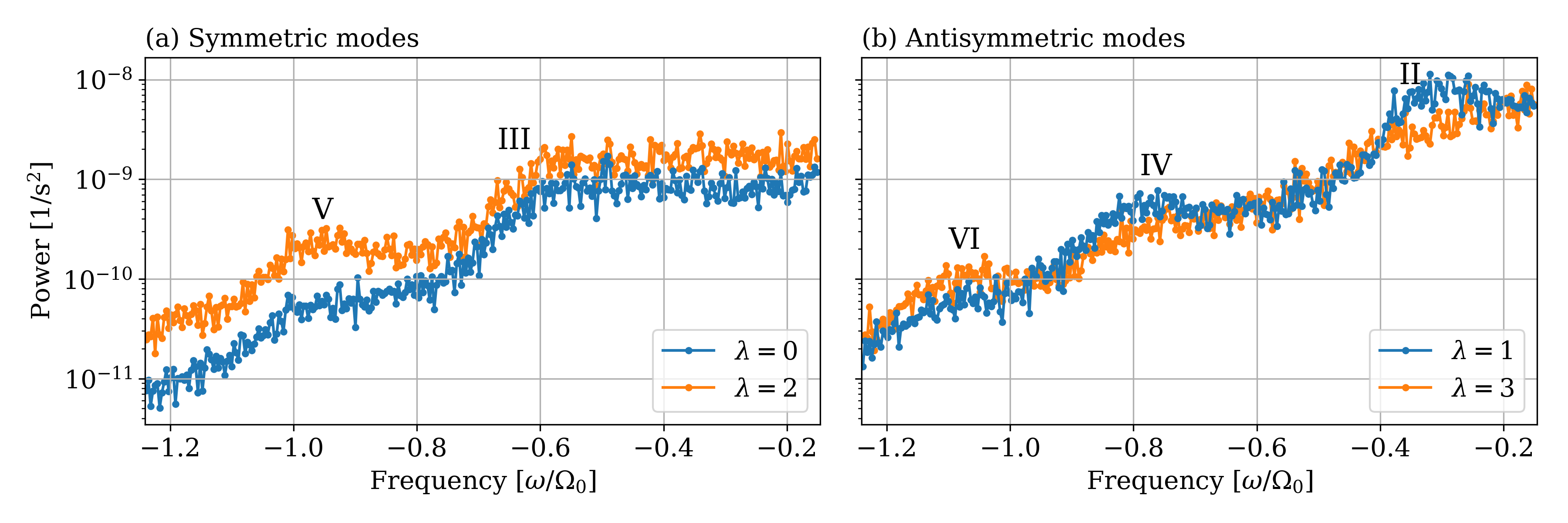}}

\vspace{-0.1in}
\caption{\normalsize  \textbf{Power profiles for the HFR modes---}Power spectra in radial vorticity for $m = 9$ at a radius of $0.935 R_{\odot}$ for latitudinal orders that are (a) symmetric ($\lambda = 0, 2$) and (b) anti-symmetric ($\lambda = 1, 3$) across the equator. The power spectrum has been binned down by a factor of 32 in frequency. We clearly see two symmetric and three antisymmetric high-frequency enhancements in power, matching the labeled ridges in Figure~\ref{fig:czwaves}. These five peaks are probably all in the same family, with higher-frequency power corresponding to modes of higher latitudinal order.
\label{fig:hfrlinecuts}
}
\end{figure*}
}

\def\hfrpoloidal{
\begin{figure*}
{\includegraphics[width = 7in, keepaspectratio=true]{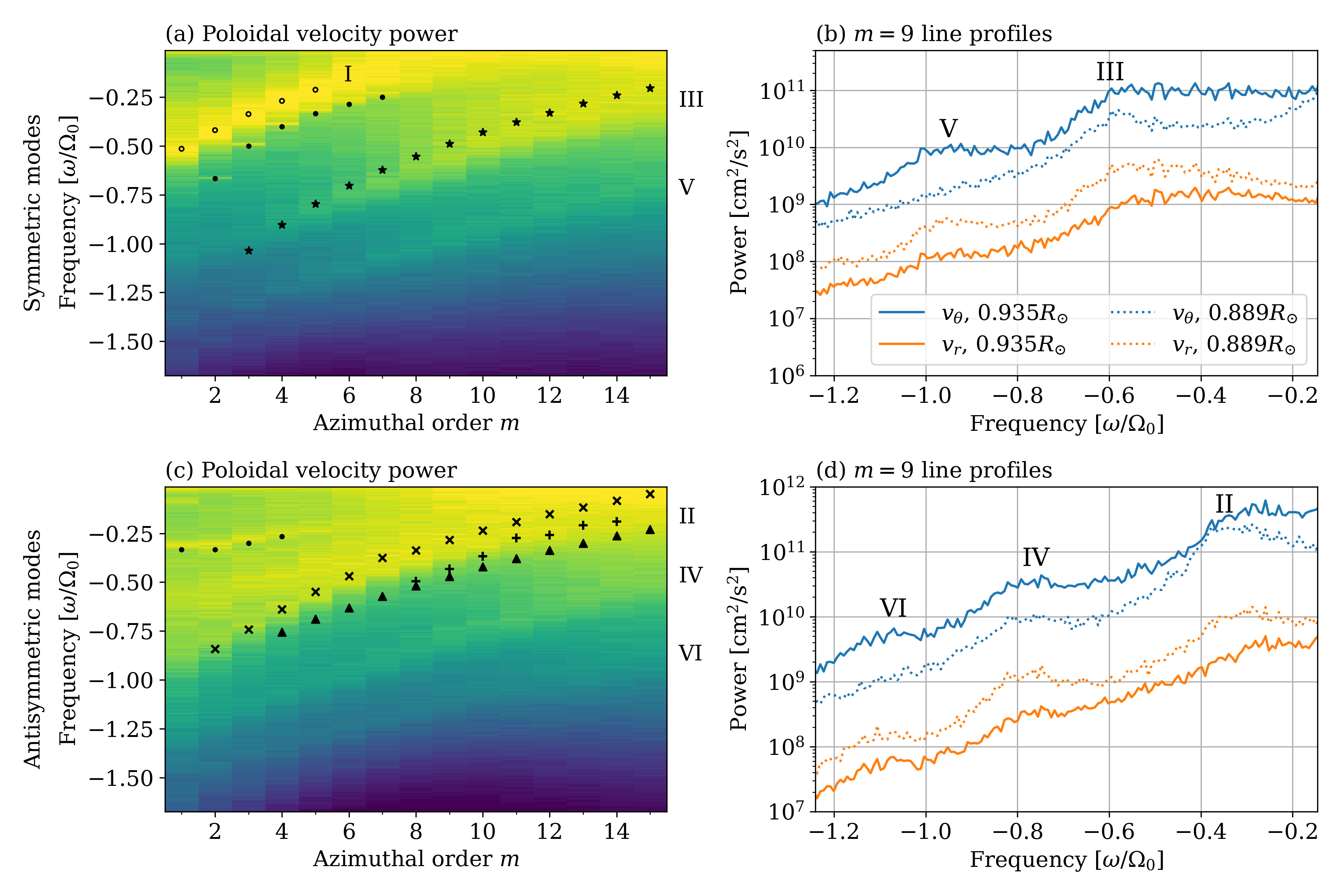}}

\vspace{-0.1in}
\caption{\normalsize  \textbf{HFR poloidal-velocity power---}(a,c) Poloidal-velocity power spectra summed over two radii near the surface, presented for (a) equatorially symmetric modes and (c) equatorially antisymmetric modes consistent with the symmetries in radial vorticity. $v_{\theta}$ shares the same symmetries as radial vorticity, but $v_r$ has the opposite. To make the ``symmetric" spectra, we thus sum over $\lambda = 0$ and 2 in $|v_{\theta}|^2$ and $\lambda = 1$ and 3 in $|v_r|^2$. For the ``antisymmetric" spectra we sum over the converse pairs. The retrograde mixed mode is again labeled I, with open circles denoting numerical values from \citet{2022AA...662A..16B}. Equatorial Rossby waves are marked with closed circles. HFR mode observations from \citet{2022NatAs...6..708H} (plus signs) and numerical eigensolver results from \cite{2023ApJS..264...21B} (symmetric modes marked in stars and antisymmetric modes marked in times signs) and \citet{2022ApJ...934L...4T} (triangles) are overplotted. Five identified HFR ridges are marked with roman numerals II--VI. (b, d) $m = 9$ power profiles for $v_{\theta}$ and $v_r$ power spectra at the two different radii, with the corresponding HFR modes marked with roman numerals. (b) shows $v_{\theta}$ summed over $\lambda = 0$ and 2 and $v_r$ summed over $\lambda = 1$ and 3. (d) shows the converse. While $v_{\theta}$ is greater at both radii, the difference between $v_{\theta}$ and $v_r$ decreases deeper in the interior.
\label{fig:hfrpoloidal}
}
\end{figure*}
}

\def\diffspectra{
\begin{figure*}
{\includegraphics[width = 7.2in, keepaspectratio=true]{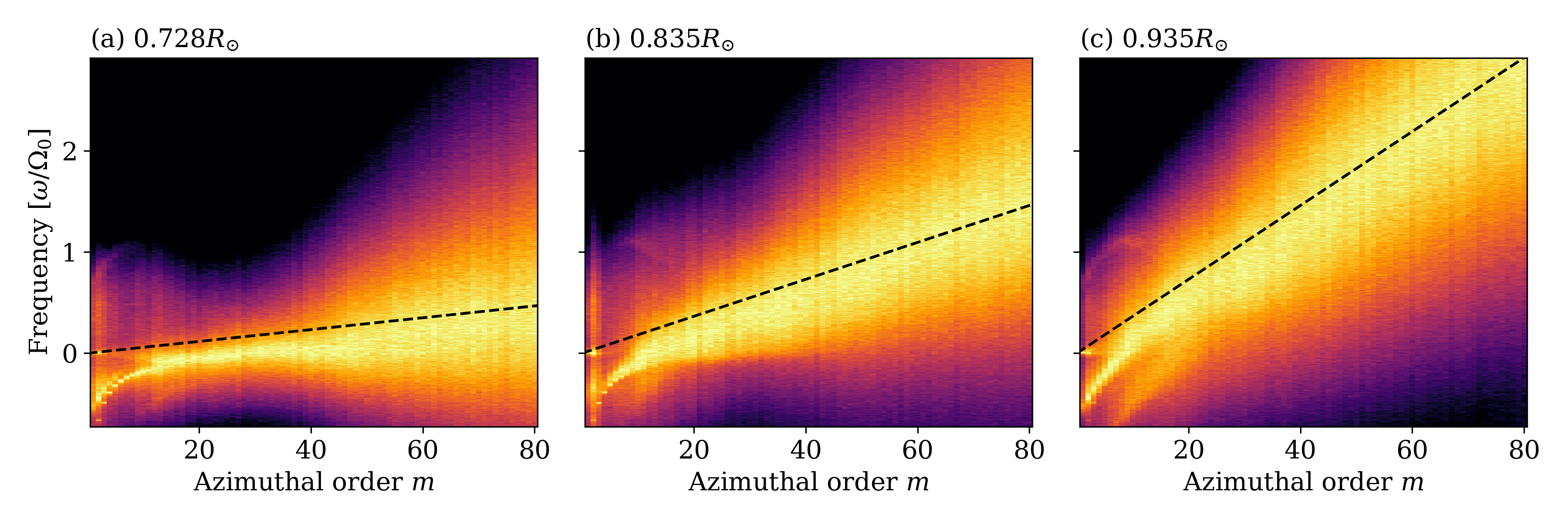}}

\vspace{-0.1in}
\caption{\normalsize  \textbf{Convective features in spectra---} Radial-vorticity spectra for $\lambda = 0$ at three distinct radii in the convection zone: (a) $0.728 \, R_{\odot}$, (b) $0.835 \, R_{\odot}$, and (c) $0.935 \, R_{\odot}$. Each column has been normalized by the largest value for that $m$ and shown on a logarithmic scale. The black dashed line shows the Doppler shift due to the radial differential rotation, given by $m \Delta \Omega$. $\Delta \Omega$ is the average differential rotation in latitude at each depth and is valued at (a) $0.007 \, \Omega_0$, (b) $0.015 \, \Omega_0$, and (c) $0.036 \, \Omega_0$. At large $m$, the convection is Doppler-shifted due to the differential rotation in the convection zone.  
\label{fig:diffspectra}
}
\end{figure*}
}

\def\radialres{
\begin{figure*}
{\includegraphics[width = 7in, keepaspectratio=true]{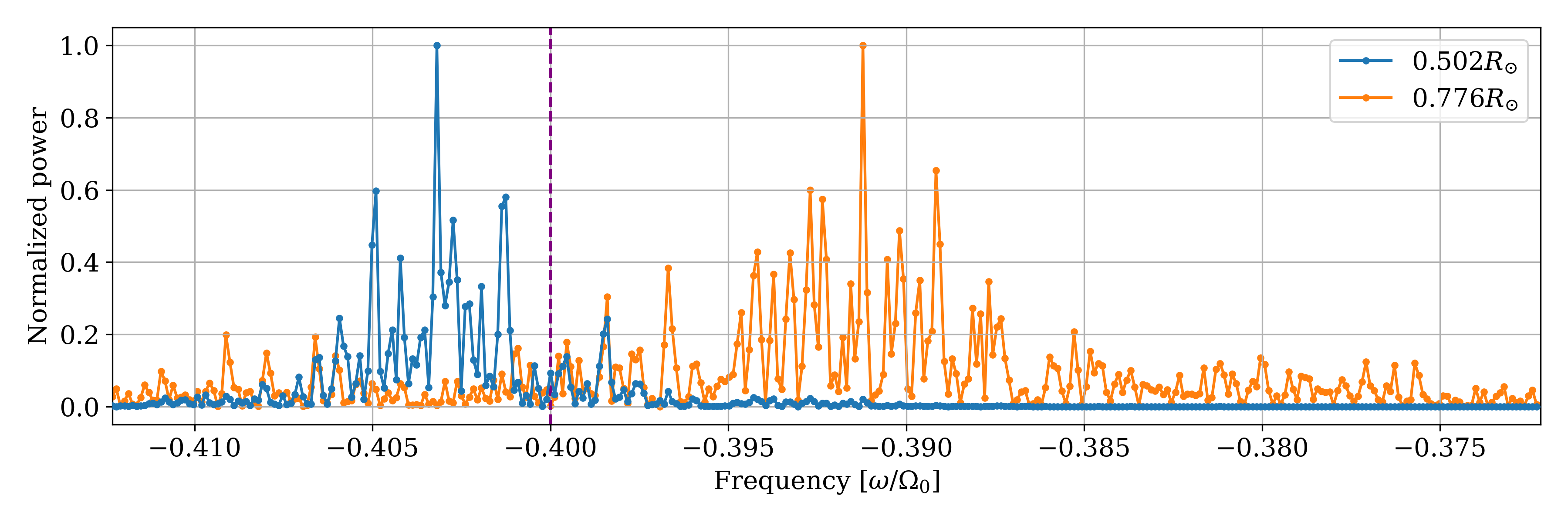}}

\vspace{-0.1in}
\caption{\normalsize  \textbf{Radial variation of the power distribution of a Rossby wave---} Line profiles for power spectra of the radial vorticity for the single spherical harmonic component $(l, m) = (4, 4)$. The blue spectrum is at a radius of $0.502 \, R_{\odot}$ located at the base of our radiative interior, and the orange is at $0.776 \, R_{\odot}$ within the lower portion of the convection zone. The purple line is the value Equation~(\ref{eq:disp}) gives when $\Delta \Omega = 0$, or what we would see if these waves lived in a region rotating at the frame rate of the simulation. We can clearly see that the equatorial Rossby waves in the radiative interior have a lower central frequency than those in the convection zone, potentially indicating two different wave families residing in two different cavities. \label{fig:radialres}
}
\end{figure*}
}

\def\freqshift{
\begin{figure}
{\includegraphics[width = 2.75in, keepaspectratio=true]{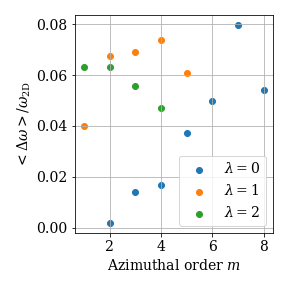}}

\vspace{-0.1in}
\caption{\normalsize \textbf{Frequency differences---}The difference in mean frequency $\langle \omega \rangle = \langle \omega \rangle_{cz} - \langle \omega \rangle_{ri}$ between modes averaged over the radiative interior and modes averaged over the convection zone. $\langle \Delta \omega \rangle$ has been non-dimensionalized by the dispersion relation value $\omega_{\rm 2D} = 2m \Omega_0/l(l+1)$. While there is not a clear relationship between latitudinal node number and frequency difference, the mean frequency for the convection zone is always less negative than those in the radiative interior. We can see that for the sectoral modes, this difference increases with $m$. 
\label{fig:freqshift}
}
\end{figure}
}

\def\mixedmodes{
\begin{figure}
{\includegraphics[width = 3in, keepaspectratio=true]{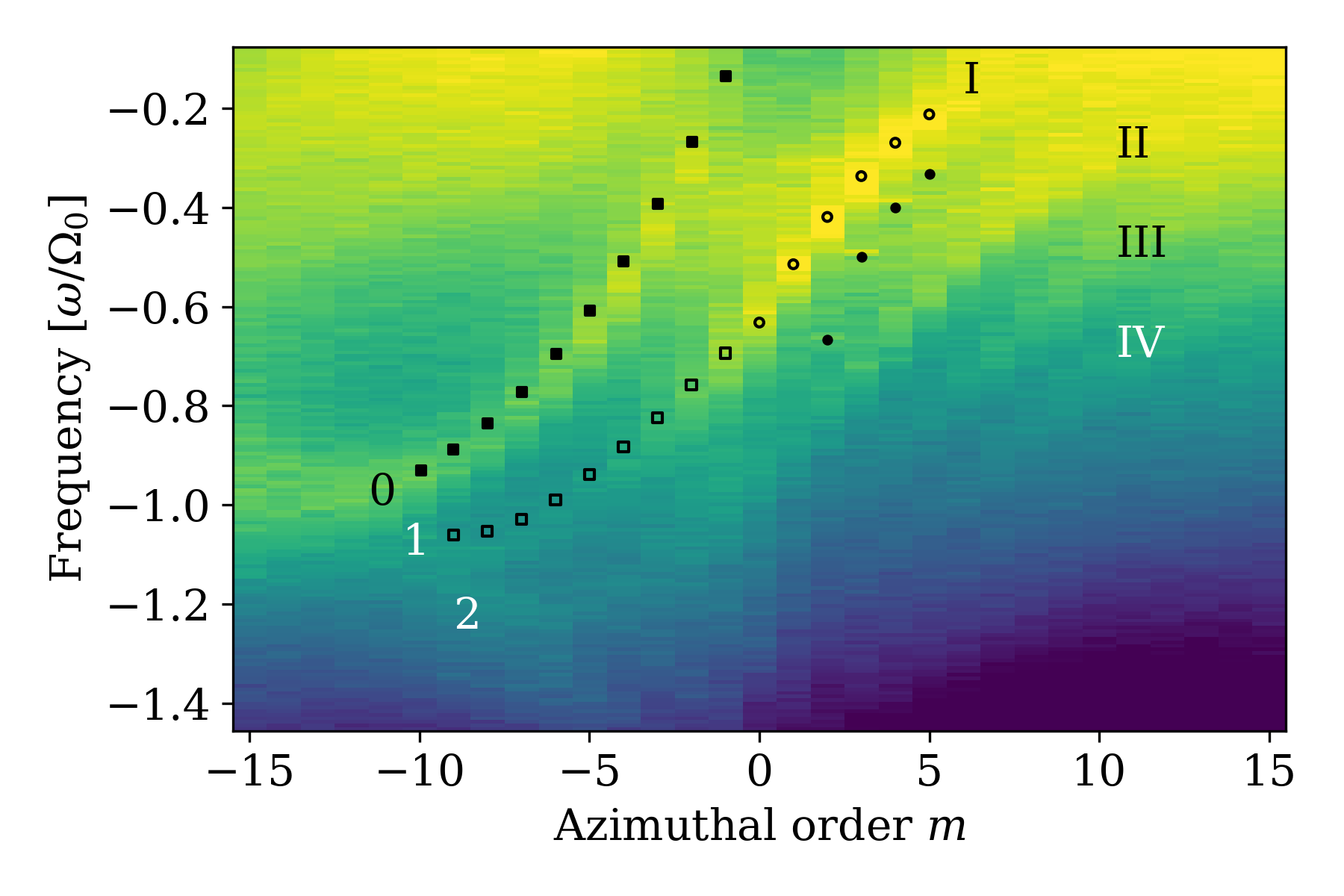}}

\vspace{-0.1in}
\caption{\normalsize \textbf{Mixed modes---}Power spectra for the poloidal velocity power, summed over $\lambda = 0$--3, averaged across two radial slices near the surface. Prograde-propagating waves have been shown with negative values of azimuthal order $m$ and negative frequencies $\omega/\Omega_0$. The Busse modes, labeled ``0", are marked with values from \citet{2022ApJ...932...68H}. Numerical values from \citet{2022AA...662A..16B} mark the previously noted mixed modes, with the prograde Roberts modes (open squares), labeled ``1", and the retrograde branch (open circles) labeled ``I". Sectoral equatorial Rossby waves are marked with filled circles. Ridge II, which corresponds with the observed HFR modes, appears to cross $m = 0$ and merge with the prograde thermal Rossby wave with two latitudinal nodes in $v_{\phi}$, labeled ``2". This potentially indicates that all of these waves are mixed modes in the same family, with thermal Rossby waves making up the prograde branch and HFR modes making up the retrograde branch.
\label{fig:mixedmodes}
}

\end{figure}
}

\def\fitting{

\begin{table*}
\centering
\textbf{Table~\ref{tab:fitting}\\}
{Thermal Rossby Wave Line-Fitting Results} 

\begin{tabular}{ccccccc}
\hline \hline

\textbf{Mode} & \textbf{Amplitude} & \textbf{Center [$\omega_0/\Omega_0$]}  & \textbf{Width [$\gamma/\Omega_0$]} & $B_0$ & $\alpha$ & \textbf{Q-factor} \\ \hline

$m = 2$ & ($2.41 \pm 0.09$) $\times 10^{-9}$ & $0.333 \pm 0.001$  & $0.033 \pm 0.002$ & ($9 \pm 7$) $\times 10^{-11}$ & ($-0.5 \pm 1$) $\times 10^{-13}$ & 10 \\ \tableline

$m = 3$ & ($2.23 \pm 0.06$) $\times 10^{-9}$ & $0.458 \pm 0.001$ & $0.050 \pm 0.002$ & ($1.2 \pm 0.4$) $\times 10^{-10}$ & ($-1.8 \pm 0.5$) $\times 10^{-13}$ & 9 \\ \tableline

$m = 4$ & ($2.67 \pm 0.07$) $\times 10^{-9}$ & $0.559 \pm 0.001$ & $0.048 \pm 0.002$ & ($2.5 \pm 0.7$) $\times 10^{-10}$ & ($-2.6 \pm 0.8$) $\times 10^{-13}$ & 12 \\ \tableline

$m = 5$ & ($2.16 \pm 0.07$) $\times 10^{-9}$ & $0.648 \pm 0.002$ & $0.067 \pm 0.005$ & ($2.5 \pm 2$) $\times 10^{-10}$& ($-3 \pm 1$) $\times 10^{-13}$ & 10 \\ \tableline

$m = 6$ & ($1.90 \pm 0.06$) $\times 10^{-9}$ & $0.741 \pm 0.003$ & $0.083 \pm 0.006$ & ($4 \pm 1$) $\times 10^{-10}$& ($-3.9 \pm 0.9$) $\times 10^{-13}$ & 9 \\ \tableline

$m = 7$ & ($2.4 \pm 0.1$) $\times 10^{-9}$ & $0.811 \pm 0.004$ & $0.077 \pm 0.008$& ($9 \pm 2$) $\times 10^{-10}$& ($-7 \pm 2$) $\times 10^{-13}$ & 10 \\ \tableline

$m = 8$ & ($3.0 \pm 0.1$) $\times 10^{-9}$& $0.879 \pm 0.003$ & $0.074 \pm 0.007$ & ($1.8 \pm 0.3$) $\times 10^{-9}$ & ($-1.5 \pm 0.2$) $\times 10^{-12}$ & 12 \\ \tableline

$m = 9$ & ($4.4 \pm 0.2$) $\times 10^{-9}$ & $0.917 \pm 0.003$ & $0.067 \pm 0.007$ & ($2.2 \pm 0.6$) $\times 10^{-9}$& ($-1.8 \pm 0.6$) $\times 10^{-12}$ & 14 \\ \tableline

$m = 10$ & ($6.9 \pm 0.2$) $\times 10^{-9}$& $0.944 \pm 0.001$ & $0.052 \pm 0.004$ & ($3.7 \pm 0.6$) $\times 10^{-9}$& ($-2.7 \pm 0.5$) $\times 10^{-12}$ & 18 \\ \tableline

\end{tabular}

\caption{Fits with the Lorentzian profile given by Equation~(\ref{eq:lorentzian}) to the power profiles for the Busse modes with $m = 2$ to $m = 10$ and $\lambda = 1$, at a radius of $0.889 R_{\odot}$. The power profiles were binned down by a factor of 32 in frequency prior to fitting.}
\label{tab:fitting}

\end{table*}
}

\def\mixedfitting{

\begin{table*}
\centering
\textbf{Table~\ref{tab:mixedfitting}\\}
{Retrograde Mixed Mode (Ridge I) Line Fitting Results} 

\begin{tabular}{ccccccc}
\hline \hline

\textbf{Mode} & \textbf{Amplitude} & \textbf{Center [$\omega_0/\Omega_0$]}  & \textbf{Width [$\gamma/\Omega_0$]} & $B_0$ & $\alpha$ & \textbf{Q-factor} \\ \hline

$m = 0$ & ($4 \pm 9$) $\times 10^{-11}$ & $-0.611 \pm 0.001$  & $0.063 \pm 0.002$ & ($2 \pm 5$) $\times 10^{-14}$ & ($2 \pm 9$) $\times 10^{-17}$ & 10  \\ \tableline

$m = 1$ & ($8 \pm 6$) $\times 10^{-11}$ & $-0.519 \pm 0.001$ & $0.053 \pm 0.002$ & ($2 \pm 3$) $\times 10^{-14}$ & ($3 \pm 4$) $\times 10^{-17}$ & 10 \\ \tableline

$m = 2$ & ($5.7 \pm 0.7$) $\times 10^{-10}$ & $-0.432 \pm 0.001$ & $0.050 \pm 0.003$ & ($8 \pm 5$) $\times 10^{-14}$ & ($1.2 \pm 0.6$) $\times 10^{-16}$ & 9 \\ \tableline

$m = 3$ & ($1.35 \pm 0.08$) $\times 10^{-9}$ & $-0.348 \pm 0.002$ & $0.051 \pm 0.005$ & ($1 \pm 1$) $\times 10^{-13}$& ($2 \pm 1$) $\times 10^{-16}$ & 7 \\ \tableline

$m = 4$ & ($2.44 \pm 0.07$) $\times 10^{-9}$ & $-0.278 \pm 0.003$ & $0.055 \pm 0.006$ & ($2.4 \pm 0.8$) $\times 10^{-13}$& ($3.5 \pm 0.7$) $\times 10^{-16}$ & 5 \\ \tableline

$m = 5$ & ($3.1 \pm 0.1$) $\times 10^{-9}$ & $-0.216 \pm 0.003$ & $0.060 \pm 0.008$& ($4 \pm 2$) $\times 10^{-13}$& ($8 \pm 2$) $\times 10^{-16}$ & 4 \\ \tableline

$m = 6$ & ($3.7 \pm 0.1$) $\times 10^{-9}$& $-0.140 \pm 0.003$ & $0.082 \pm 0.006$ & ($3 \pm 2$) $\times 10^{-13}$ & ($9 \pm 2$) $\times 10^{-16}$ & 2 \\ \tableline

$m = 7$ & ($3.1 \pm 0.2$) $\times 10^{-9}$ & $-0.100 \pm 0.003$ & $0.060 \pm 0.007$ & ($1.5 \pm 0.5$) $\times 10^{-12}$& ($3.7 \pm 0.4$) $\times 10^{-15}$ & 2 \\ \tableline

\end{tabular}

\caption{Lorentzian fits (Equation~(\ref{eq:lorentzian})) to the power spectra of the retrograde mixed modes (ridge I) for $l=m=1$ to $l=m=7$, along with the $(l, m) = (2, 0)$ mode, at a radius of $0.935 R_{\odot}$. Power profiles were binned down by a factor of 32 in frequency prior to fitting.}
\label{tab:mixedfitting}

\end{table*}
}

\def\summarytable{
\begin{table*}
\centering
\textbf{Table~\ref{tab:summarytable}\\}
{Summary of Observed and Modeled Solar Inertial Modes} 

\begin{tabular}{p{0.35\textwidth}>{\centering}p{0.15\textwidth}>{\centering}p{0.19\textwidth}>{\centering\arraybackslash}p{0.19\textwidth}}
\hline \hline
\textbf{Type} & \textbf{Simulation Location} & \textbf{Have they been observed and identified?}  & \textbf{Have they been identified in other models?} \\ \hline
Equatorial Rossby waves, sectoral modes & RI and CZ & Yes\textsuperscript{e.g., 1} & Yes\textsuperscript{e.g., 2,3}      \\ \tableline 
Equatorial Rossby waves, tesseral modes & RI and CZ & Maybe\textsuperscript{4} & Yes\textsuperscript{4,5}  \\ \tableline
Retrograde mixed modes & CZ & Maybe\textsuperscript{6} & Yes\textsuperscript{2}   \\ \tableline
Thermal Rossby waves & CZ & No & Yes\textsuperscript{e.g., 2, 7, 8}   \\ \tableline
HFR modes, equatorially symmetric $\lambda = 0$ & CZ & No & Yes\textsuperscript{9}    \\ \tableline
HFR modes, equatorially anti-symmetric $\lambda = 1$ & CZ & Yes\textsuperscript{10} & Yes\textsuperscript{4,9,11}    \\ \tableline
HFR modes, $\lambda \geq 2$ & CZ & No & No  \\   \tableline
High-latitude modes & -- & Yes\textsuperscript{12} & Yes\textsuperscript{2,5}  \\\tableline
Critical-latitude modes & -- & Yes\textsuperscript{12} & Yes\textsuperscript{2,3,5}
\end{tabular}

\textsuperscript{1} \citet{2018NatAs...2..568L}, \textsuperscript{2} \citet{2022AA...662A..16B,2022AA...666A.135B}, \textsuperscript{3} \citet{2020AA...642A.178G}, \textsuperscript{4} \citet{2022ApJ...934L...4T}, \textsuperscript{5} \citet{2022AA...664A...6F}, \textsuperscript{6} \citet{Waidele2023},  
\textsuperscript{7} \citet{2022ApJ...932...68H}, \textsuperscript{8} \citet{2023ApJ...943..127H}, \textsuperscript{9} \citet{2023ApJS..264...21B}, \textsuperscript{10} \citet{2022NatAs...6..708H}, \textsuperscript{11} \citet{2023arXiv231110414B}, \textsuperscript{12} \citet{2021AA...652L...6G}

\caption{A summary of inertial waves, where they are found in this simulation---radiative interior (RI) and/or convection zone (CZ)---and the observations and models that have referred to them in the recent literature.}
\label{tab:summarytable}
\end{table*}
}

%% file: S1-Introduction.tex
\section{Introduction}
\label{sec:intro}

Although Rossby waves and other inertial oscillations have been extensively studied for the past eighty years, their recent observation on the Sun has revived an interest in the solar physics community, particularly in their potential uses for helioseismology. The Rossby waves themselves are likely to be a sensitive diagnostic of the convection zone’s superadiabaticity \citep[e.g.,][]{Gilman1987}, and the subsequent detection of high-latitude inertial oscillations \citep{2021AA...652L...6G} has the potential to open up seismic exploration of the polar caps where many dynamo processes are likely to operate.

Rossby waves were first observed on the surface of the Sun by \citet{2018NatAs...2..568L}, who observed sectoral modes, i.e., modes with eigenfunctions that lack latitudinal nodes. This initial discovery has since been confirmed using a variety of techniques and data sets \citep[e.g.,][]{Alshehhi2019, Hanasoge2019, 2019A&A...626A...3L, Hanson2020, 2020A&A...634A..44P, 2021AA...652L...6G, Hathaway2021, Waidele2023}.
More recently, \citet{2022ApJ...934L...4T} noted the possible presence in the observational data of tesseral Rossby modes, which have one or more latitudinal nodes in their eigenfunctions. Other types of inertial modes have also been observed, such as high-latitude and critical-latitude inertial modes \citep{2021AA...652L...6G} and the mysterious high-frequency retrograde (HFR) waves \citep{2022NatAs...6..708H}, which currently lack a thorough theoretical explanation.

Conversely, there are theoretically expected inertial oscillations that have not yet been observed. For example, thermal Rossby waves are a class of prograde-propagating inertial waves that have long appeared in simulations \citep[e.g.,][]{1970JFM....44..441B, Hindman2020, 2022AA...666A.135B, 2022ApJ...932...68H, 2023ApJ...943..127H, 2023ApJ...958...48J} and in laboratory experiments \citep[e.g.,][]{Mason1975, Busse1982, Azouni1985, Chamberlain1986, Cordero1992, Smith2014, Lonner2022}, but have yet to be detected observationally in the Sun. 

\citet{2022AA...662A..16B} additionally predicts the presence of a ``mixed mode" comprised of a prograde branch of thermal Rossby waves with one latitudinal node and a retrograde branch of inertial waves with one radial node. This mode is mixed in the sense that the Yanai mode is mixed \citep[e.g.][]{Gill1982}, with differing behavior between the retrograde and prograde branches. The prograde branch of Yanai waves is essentially composed of internal gravity waves, while the retrograde branch is composed of planetary Rossby waves. \citet{2022AA...662A..16B} refers to the retrograde branch of the mixed mode that they have identified as an equatorial Rossby wave with one radial node. However, this mode has significant vertical motion and thus does not behave like a typical equatorial Rossby wave. For this reason, we refer to them as retrograde mixed modes throughout. In this work, we additionally suggest that this mode is just a single member of a family of mixed modes that reside only in convection zones where the stratification is nearly neutrally stable. This family may include the observed HFR modes, which also evince strong vertical motions. 

Although Rossby waves have been a mainstay in meteorology since their discovery in the late 1930s \citep{Rossby1939RelationBV, 1940TrAGU..21..262H}, the theoretical work on Rossby waves in a stellar context has been more sporadic. \citet{1978MNRAS.182..423P} introduced the concept of ``r-modes" to the stellar physics community when they noticed Rossby waves amongst the solution set for low-frequency oscillations of rotating stars. \citet{1981A&A....94..126P}, \citet{1981Ap&SS..78..483S}, and \citet{1982ApJ...256..717S} studied the eigenfunctions and radial structure of these modes in the case of slow, spatially uniform rotation. \citet{1986SoPh..105....1W} focused on the radial behavior of r-modes in the Sun's interior, concluding that the r-modes exist in two separate cavities: one cavity in the radiative interior and one in the convection zone. The splitting of the Rossby waves into two distinct families arises due to the sign of the Ledoux discriminant; waves in an unstable stratification propagate in a fundamentally different manner than those in a stable stratification \citep[see also][]{2023A&A...671A..91A}.

The recent observations of inertial waves in the Sun have prompted a flurry of theoretical work to better understand which modes can exist in the convection zone and how different physical processes impact these modes. The bulk of this work has been in the form of linear eigenmode calculations \citep[e.g.,][]{2020AA...642A.178G, 2022AA...662A..16B, 2022ApJ...934L...4T, 2022ApJ...932...68H, 2023ApJ...943..127H, 2023ApJS..264...21B, 2023arXiv231110414B}, but the work of \citet{2022AA...666A.135B} examined the inertial modes appearing in a nonlinear numerical simulation with a spatial domain that spanned the solar convection zone. Here, we adopt a similar methodology and explore the inertial modes that manifest in a numerical simulation, but our spatial domain includes both a convection zone and a significant portion of the underlying stably stratified radiative interior.

We find a veritable menagerie of waves that have all been self-consistently excited in a solar-like environment. In particular, the model features equatorial Rossby waves in the radiative interior, tesseral equatorial Rossby waves in the convection zone, and high-frequency retrograde (HFR) waves of latitudinal orders that have not been noted previously.

The paper is organized as follows. Section~\ref{sec:methods} provides details of the simulation and  code. Section~\ref{sec:summary} describes the spectra that make up the bulk of our analysis. Section~\ref{sec:RIWaves} focuses on the equatorial Rossby waves found in the radiative interior, while Section~\ref{sec:CZWaves} covers the many types of waves found in the convection zone. Section~\ref{sec:discussion} summarizes our findings and discusses the implications of our work, including the possibility of two families of Rossby waves as well as the classification of HFR modes. 

%% file: S2-Simulation.tex
\section{Methods}
\label{sec:methods}

\subsection{The Rayleigh Code}
This numerical model was generated using the {\it Rayleigh} convection code \citep{2021zndo...5774039F, 2016ApJ...818...32F, 2016GGG....17.1586M}, which solves the fluid equations in rotating spherical geometry using the pseudo-spectral algorithm from \citet{1984JCoPh..55..461G} and \citet{Clune1999}. Each physical variable is represented as a linear combination of spectral components, with each component consisting of the product of a spherical harmonic ${Y}_{l}^{m}(\theta ,\phi )$ and a Chebyshev polynomial $T_k(r)$. The three spherical coordinates $r$, $\theta$, and $\phi$, are the radius, colatitude, and longitude, respectively, and $\unitv{r}$, $\unitv{\theta }$, and $\unitv{\phi}$ represent the local unit vectors pointing in these three directions. In the spectral representation, $l$ is the harmonic degree, $m$ is the azimuthal order, and $k$ is the radial order. The fluid variables are updated at each timestep using a second-order Crank–Nicolson scheme that handles most of the linear terms in the fluid equations directly in spectral space. The nonlinear terms and the Coriolis term are calculated using an Adams–Bashforth algorithm that is performed in physical space after transformation from the spectral representation.

Because our flows are low Mach number, we linearize the thermodynamic variables about a reference state and assume a solenoidal mass flux. Specifically, we use the Lantz-Braginsky-Roberts (LBR) formulation \citep{Lantz1992,BR1995} of the anelastic MHD equations \citep[e.g.,][]{TheAnelasticApproximationforThermalConvection,1981ApJS...45..335G} in the rotating frame, given by
\begin{multline}
    \hat{\rho} \left[ \pdv{\vb*{v}}{t} + \vb*{v} \cdot \grad \vb*{v} + 2 \bvec{\Omega}_0 \cross \vb*{v} \right] = \frac{g\hat{\rho}\,s^\prime}{c_P}  \unitv{r} \\- \hat{\rho} \grad \left( \frac{P^\prime}{\hat{\rho}} \right) + \frac{1}{4 \pi} \left(\curl \vb*{B} \right) \cross \vb*{B} + \div{\mathcal{D}} \; ,
\end{multline}
\begin{multline}
    \hat{\rho} \hat{T} \left[ \pdv{s^\prime}{t} + \vb*{v} \cdot \grad s^\prime + v_r \dv{\hat{s}}{r}\right] = \div \left[\hat{\rho} \hat{T} \kappa \grad s^\prime \right] \\+ Q + \Phi + \frac{\eta}{4 \pi} \left|\curl \vb*{B} \right|^2 \; ,
\end{multline}
\begin{align}
    & \pdv{\vb*{B}}{t} = \curl \left(\vb*{v} \cross \vb*{B} - \eta \curl \vb*{B} \right) \; , \\
    & \mathcal{D}_{ij} = 2 \hat{\rho} \nu  \left[e_{ij} - \frac{1}{3} \left( \div \vb*{v} \right) \delta_{ij} \right] \; , \\
    & \Phi = 2 \hat{\rho} \nu \left[e_{ij}e_{ij} - \frac{1}{3} \left(\div \vb*{v} \right)^2 \right] \; , \\
    & \div \left(\hat{\rho} \vb*{v} \right) = 0  \; ,\\
    & \div \vb*{B} = 0  \; ,
\end{align}
\noindent where the primary variables are the velocity $\vb*{v}$, magnetic field $\vb*{B}$, and the fluctuations of the gas pressure $P^\prime$ and specific entropy density $s^\prime$ about the corresponding reference-state profiles. The angular velocity of the rotating reference frame, $\bvec{\Omega}_0$, is aligned with the axis of the spherical coordinate system. The gravitational acceleration is given by $g(r)$; the viscous, thermal and magnetic diffusivities by $\nu(r)$, $\kappa(r)$, and $\eta(r)$, respectively; $c_P$ is the specific heat capacity at constant pressure; and $e_{ij}$ is the rate-of-strain tensor. The radial profiles $\hat{\rho}(r)$ and $\hat{T}(r)$ are the time-independent, spherically symmetric reference-state density and temperature, which satisfy the ideal gas law, and $\text{d} \hat{s}/ \text{d} r$ is the reference-state entropy gradient. The equation of state for an ideal gas is linearized and expressed in terms of the fluctuations about this reference state. The momentum equation includes the Coriolis, buoyancy, pressure gradient, Lorentz, and viscous forces. The thermal energy equation includes conductive, Ohmic, and viscous heating, as well as an internal heating term $Q(r)$ that represents radiative heating. The internal heating is normalized to deposit a solar luminosity of heat, which is distributed preferentially in roughly the bottom third of the convection zone \citep[for details, see][]{2016ApJ...818...32F}.

\subsection{The Numerical Experiment}

This MHD model evinces a self-consistently established tachocline that is in a statistical steady state. We previously explored the properties of this tachocline and the balances that enable its formation in \citet{Matilsky2022} \citep[see also][]{Matilsky2023}. The simulation spans a spherical shell covering the upper radiative interior and lower convection zone, extending between the radii $r_{\text{min}} = 0.491 \, R_{\odot}$ and $r_{\text{max}} = 0.947 \, R _{\odot}$, where $R_{\odot} = 6.957 \cross 10^{10}$ cm is the solar radius. The transition from convective stability to instability is primarily controlled by the entropy gradient of the reference state $\text{d} \hat{s} / \text{d} r$ and occurs at $r_{\text{bcz}} = 0.729 \,  R_{\odot}$. Convective downflows overshoot into a thin layer within the stable region, the base of which is $r_{\text{ov}} = 0.710 \, R_{\odot}$.

The spherical shell spans roughly the upper two density scale heights of the radiative interior and the bottom three density scale heights of the Sun's convection zone. The horizontal grid resolution is $N_{\theta} = 384$ and $N_{\phi} = 768$, corresponding to a maximum spherical harmonic degree after de-aliasing of $l_{\text{max}} = 255$. In the radial direction, there are three stacked Chebyshev domains with 64 points each, with interior boundaries located at $0.669~R_\odot$ and $0.719~R_\odot$. The stacked grids provide substantially higher radial resolution in the tachocline and overshoot layer \citep{Matilsky2022}.

 The background entropy gradient is a positive constant in the radiative interior and zero in the convection zone. The following smooth profile connects the two domains:
 \begin{align}
     \dv{\hat{s}}{r} = \sigma \begin{cases}
       1 & r \leq r_0 - \delta \; , \\
        1 - \left[1 - \left(\frac{r - r_0}{\delta}\right)^2 \right]^2  & r_0 - \delta < r < r_0 \; ,\\
        0 & r \geq r_0 \; ,
     \end{cases}
 \end{align}
 
\noindent where $\sigma \equiv 10^{-2}$ erg g$^{-1}$ K$^{-1}$ cm$^{-1}$, $\delta \equiv 0.05 \,R_{\odot}$, and $r_0 = 0.719 \,R_{\odot}$. The gravitational acceleration is given by
 \begin{align}
     g(r) = \frac{G M_{\odot}}{r^2},
 \end{align}

\noindent where $G$ is the universal gravitational constant and $M_{\odot} = 1.989 \times 10^{33}$ is the solar mass.

The rotation rate is given by $\Omega_0/2\pi = 1370$ nHz, and a solar luminosity $L_{\odot} = 3.85 \cross 10^{33}$ erg s$^{-1}$ is driven through the convection zone via the temporally steady internal-heating profile $Q(r)$ \citep[for details see][]{Matilsky2022}. The viscous, thermal, and magnetic diffusivities at the top of the domain are given by $\nu(r_{\max}) = \kappa(r_{\max}) = 5 \cross 10^{12}$ cm$^2$ s$^{-1}$ and $\eta(r_{\max}) = 1.25 \cross 10^{12}$ cm$^2$ s$^{-1}$, respectively. All diffusivities increase with radius, varying like $\sim\hat{\rho}^{-1/2}$.

At both boundaries, we use stress-free and impenetrable conditions on the velocity, i.e.,
\begin{equation}
    v_r = \pdv{r} (v_{\theta}/r) = \pdv{r} (v_{\phi}/r) = 0 \; .    
\end{equation}

\noindent We impose a fixed conductive flux at the inner and outer spherical boundaries, using
\begin{eqnarray}
    \left.\pdv{s^\prime}{r}\right|_{r_{\rm min}} &=& 0 \; ,
\\
    \left.\quad \pdv{s^\prime}{r}\right|_{r_{\rm max}} &=& \left.-\frac{L_{\odot}}{4 \pi r^2 \kappa(r) \hat{\rho}(r) \hat{T}(r)} \right|_{r_{\rm max}}  \; .
\end{eqnarray}
    
\noindent No heat is allowed to enter at the bottom of the domain, and at the top of the domain, heat exits via thermal conduction. Potential field matching conditions are used on the magnetic fields at both boundaries.  

\renewcommand{\arraystretch}{1.5}
\begin{table}[t]
\centering
\textbf{Table~\ref{tab:parameters}\\}
{Non-dimensional Fluid Parameters}
\begin{tabular}{c c c}
\hline \hline

\textbf{Parameter} & \textbf{Definition} & \textbf{Value} \\ \hline
    $\rm Ra_F$ & \large{$\frac{\tilde{g} \tilde{F} H^4}{c_P \tilde{\rho} \tilde{T} \tilde{\nu} \tilde{\kappa}^2}$} & $5.68 \cross 10^5$ \\
    $\rm Ek$ & \large{$\frac{\tilde{\nu}}{2 \Omega_0 H^2}$} & $5.35 \cross 10^{-4}$ \\
    $\rm Pr$ & \large{$\frac{\tilde{\nu}}{\tilde{\kappa}}$} & $1$ \\
    $\rm Pr_m$ & \large{$\frac{\tilde{\nu}}{\tilde{\eta}}$} & $4$ \\
    $\rm Ro_c$ & \large{$\left(\frac{\text{Ra}_{\text{F}} \, \text{Ek}^2}{\text{Pr}}\right)^{\frac{1}{2}}$} & 0.403 \\
    $\rm Bu$ & \large{$\frac{\tilde{N}^2}{4 \Omega_0^2}$} & $6.35 \cross 10^{3}$
\end{tabular}
\caption{The flux Rayleigh number $\rm Ra_F$, Ekman number Ek, Prandtl number Pr, magnetic Prandtl number $\rm Pr_m$, convective Rossby number $\rm Ro_c$, and buoyancy parameter Bu for the numerical simulation have been defined using volumetric averages of the reference state profiles.  Averages for the first five parameters are taken over the convection zone only, while the buoyancy frequency is averaged over the radiative interior only. These quantities are indicated by use of a tilde over the variable.}
\label{tab:parameters}
\end{table}

Although we ran the simulation dimensionally, its parameter regime can be uniquely described by the non-dimensional parameters given in Table \ref{tab:parameters}, where the tildes indicate volume-averaged quantities and $H=r_{\rm max} - r_{\rm bcz}$ is the depth of the convection zone. We adopt the flux Rayleigh number Ra$_{\text{F}}$, where $F$ is the energy flux imposed by radiative heating, i.e., the flux not carried by the radiation field \citep[see ][]{2016ApJ...830L..15F}. Ek is the Ekman number, which controls the importance of viscosity relative to rotation. The convective Rossby number Ro$\rm _c$ characterizes the relative strength of the buoyancy and Coriolis forces and often determines the degree of rotational influence on the convection \citep{2020PhRvR...2d3115A,2022ApJ...938...65C}. The Prandtl number Pr and the magnetic Prandtl number $\rm Pr_m$ specify the various ratios of the viscous, thermal, and magnetic diffusion coefficients. The buoyancy parameter Bu assesses the stiffness of the radiative interior.

\subsection{Simulation Summary}
\dr
\simparams

This model has a solar-like differential rotation profile, with the equator rotating faster than the polar regions, illustrated in Figure~\ref{fig:diff_profile}. Dynamo action creates a large-scale, cycling, non-axisymmetric magnetic field whose torque enforces solid-body rotation in the radiative interior and maintains a tachocline \citep[see][]{Matilsky2022}. The rotation rate is given by $\Omega(r) = \Omega_0 + \langle v_{\phi} \rangle/r \sin \theta$, where the angular brackets refer to a combined temporal and zonal average, and $\Omega_0$ is the frame rate. Figure  ~\ref{fig:diff_profile} shows non-dimensionalized differential rotation profile $(\Omega - \Omega_0)/\Omega_0$. Here and  in Figure ~\ref{fig:simparams}, the pink dash-dotted lines indicate the base of the tachocline at $r_{\text{tach}} = 0.641 \, R_{\odot}$ and the black dashed lines indicate the base of the convection zone at $0.729 \, R_{\odot}$. The model evinces $0.04 \, \Omega_0$ difference in the rotation rate from pole to equator in the convection zone, whereas in the radiative interior the pole-to-equator contrast is roughly $0.0015 \, \Omega_0$.

Figures~\ref{fig:simparams}a and \ref{fig:simparams}b display the kinetic and magnetic energy densities as functions of radius across the entire domain. The contributions to the kinetic energy from the horizontal and radial velocity components are similar in magnitude throughout the convection zone. Across the tachocline, the contribution from the radial velocity component drops by eight orders of magnitude, while the horizontal velocities only decrease by four orders of magnitude. The fact that horizontal velocities tend to be significantly higher than vertical velocities in simulations has been noted multiple times in past work  \citep[e.g.,][]{2015A&A...581A.112A,2015ApJ...813...95L}. In Section \ref{sec:RIWaves}, we determine that this enhanced horizontal power is due to a rich spectrum of equatorial Rossby waves.  Similarly, the magnetic field is primarily horizontal in the radiative interior compared to the convection zone, with the radial field component being smaller by roughly an order of magnitude (or two orders of magnitude in the square of the field components).

Figure~\ref{fig:simparams}c, shows the radial behavior of the dynamic Elsasser number as defined by \citet{2012E&PSL.333....9S}:
\begin{equation} \label{eqn:Elsasser}
    \Lambda_d(r) = \frac{1}{4 \pi}\frac{B^2}{2 \hat{\rho} \Omega_0 U L},
\end{equation}

\noindent where $B(r)$ is the spherically averaged rms field strength, $L$ is the characteristic length scale for the magnetic field, and $U(r)$ is the spherically averaged rms flow speed. We choose $L = \pi R_\odot/2$, or a length scale corresponding to a large-scale field component with a harmonic degree of $l = 2$. \citet{Matilsky2023} has demonstrated that the magnetic field in this simulation is concentrated in the low-order spherical-harmonic components. The Elsasser number compares the strength of the Coriolis force to that of the Lorentz force. An Elsasser number greater than one indicates that the Lorentz force dominates the dynamics, whereas an Elsasser number of less than one indicates that the Coriolis force dominates. The Elsasser number in this simulation is always quite small ($\Lambda_d < 10^{-2}$), hence the wave dynamics are rotationally rather than magnetically constrained. Thus, even though the magnetic field is crucial in the dynamics of the mean flows, the inertial waves can essentially be treated as nonmagnetic and hydrodynamic.

%% file: S3-SpectraSummary.tex
\section{Summary of Wave Spectra}
\label{sec:summary}

We illustrate the wave fields in our model through spectra of the radial component of the vorticity. These spectra are obtained with Fourier transforms in time $t$ and spherical harmonic transforms in the horizontal coordinates $\theta$ and $\phi$. The resulting decomposition of the data is four-dimensional with each spectral component being a function of temporal frequency $\omega$, radius $r$, harmonic degree $l$, and azimuthal order $m$.  An individual spherical harmonic possesses a number of nodes in latitude equal to $\lambda = l - |m|$, which we call the latitudinal order. The waves with $\lambda = 0$, or $l = \pm m$, are the sectoral modes, which have zero nodes in latitude. The tesseral modes have $\lambda > 0$ (or $l > |m|$) and have at least one latitudinal node.  Without loss of generality, we often choose to illustrate only positive values of the azimuthal order ($m>0$) and adopt the convention that negative frequencies correspond to modes propagating in the retrograde direction and positive frequencies in the prograde direction.

We generate spectra at fifteen different radii throughout the spatial domain, with five samples each in the radiative interior (0.502 $R_{\odot}$--0.65 $R_{\odot}$), tachocline (0.678 $R_{\odot}$--0.728 $R_{\odot}$), and convection zone (0.731 $R_{\odot}$--0.935 $R_{\odot}$). The simulation was run with a maximum spherical harmonic degree of 255, but we only present spectra for $m \leq 80$. The Fourier transform was taken over a single contiguous realization of about 200 years in duration (or 8770 rotational periods), resulting in spectra with a Nyquist frequency of 5000 nHz and a frequency resolution of 0.156 nHz.  In terms of a dimensionless frequency $\omega/\Omega_0$, this corresponds to a Nyquist frequency of 3.6 and frequency resolution of $1.1 \times 10^{-4}$.

\spectrasummary

Figure~\ref{fig:Wave spectra summary} displays power spectra of the radial vorticity, averaged over the radiative interior (left), tachocline (center), and convection zone (right). To more clearly see each type of wave and its behavior, the top row displays the sum of wave modes with even $\lambda = 0, 2, 4,$ and $6$, which are symmetric across the equator, while the bottom row displays the sum of modes with odd $\lambda = 1, 3,$ and $5$, which are anti-symmetric across the equator. Summing the power only over large spatial scales, or low latitudinal orders ($\lambda < 7$), removes a significant amount of nonmodal power arising from convective noise. We additionally scale each $m$ individually to make the mode relationships more visible across all length scales. Because the dynamic range of each plot was chosen to maximize the visibility of select features, we have elected not to display a color bar. Instead, corresponding line profiles for each wave type provide the relevant magnitudes. The resulting spectrum clearly illustrates in a single image the various types of inertial oscillations that are present in each portion of the spatial domain. 

We discuss each of these wave types in the following sections. In the radiative interior, we find both sectoral and tesseral modes of equatorial Rossby waves (Section~\ref{sec:RIWaves}). In the tachocline and convection zone, sectoral and tesseral Rossby waves once again appear (Section \ref{sec:czrw}), along with thermal Rossby waves (Section \ref{sec:thermal}), retrograde mixed modes (Section \ref{sec:czrw,n1}), and HFR modes with what are potentially their latitudinal overtones (Section \ref{sec:hfr}).  

\subsection{Doppler-Shift}

If the modes live in a region rotating at a rate $\Omega$ different from our frame rate $\Omega_0$ (i.e., $\Omega = \Omega_0 + \Delta\Omega$), then a wave of azimuthal order $m$ and intrinsic frequency $\omega'$ will be observed in the reference frame of the simulation (rotating at rate $\Omega_0$) as having the Doppler-shifted frequency

\begin{equation} \label{eq:doppler}
    \omega = \omega' + m \Delta \Omega \, ,
\end{equation}

\noindent \citep[e.g.,][]{1968RSPSA.302..529B}. 

It is important to note that both the observed and intrinsic frequency of a normal mode do not change with spatial position; the same $\Delta \Omega$ correction can be applied across the entire domain. The Doppler shift that relates these two frequencies is a spatial average of the rotation-rate difference over the region where the mode's eigenfunction has a significant amplitude. Therefore, the Doppler shift depends on the radial and latitudinal eigenfunctions. For waves living in the radiative interior, which is rotating mostly like a solid body, we can easily apply a $\Delta \Omega$ that is simply the difference between our frame rate and the rotation rate of the radiative interior. Since the radiative interior is rotating slightly slower than our frame rate, the correction is negative ($\Delta\Omega < 0$). For waves that reside in the convection zone, the situation is more complicated because the rotation profile of the convection zone varies significantly in both radius and latitude. Different modes that occupy different parts of the convection zone will sense a different mean rotation rate. Thus, without knowing the detailed shape of the eigenfunctions, we can only make informed guesses for the exact Doppler shift that should be applied.

%% file: S4-RIWaves.tex
\section{Waves in the Radiative Interior}
\label{sec:RIWaves}

In the radiative interior, we observe a rich spectrum of equatorial Rossby waves. Both sectoral and tesseral modes are present, with power distributed relatively widely amongst the distinct modes. We explore the nature of these waves in the following subsections.

\vspace{0.8cm}
\subsection{Rossby-Wave Spectra}

The left-hand column of Figure \ref{fig:Wave spectra summary} makes clear that both sectoral and tesseral equatorial Rossby waves are present in the radiative interior of our simulation. Figure \ref{fig:radispectra}a displays the power spectrum in radial vorticity, averaged over the radiative interior, summed over all latitudinal orders (or equivalently, over all harmonic degrees). The black dots denote a theoretical estimate of the frequencies,

\begin{equation}
    \label{eq:disp}
    \omega = m \Delta \Omega - \frac{2 m \Omega_0}{l(l+1)}.
\end{equation}

\noindent This dispersion relation is derived for purely hydrodynamic, equatorial Rossby waves in two horizontal dimensions with solid-body rotation at a frame rate of $\Omega_0$ \citep{1940TrAGU..21..262H}. For such a spherically symmetric system, the eigenfunctions for the Rossby waves are pure spherical harmonics. Since the radiative interior of our model is rotating slightly slower than the frame rate (with a difference of $\Delta\Omega\approx -0.0015 \, \Omega_0$, see Figure \ref{fig:diff_profile}), we have applied a Doppler shift to the traditional dispersion relation.

\rwradi

In Figure \ref{fig:radispectra}a, the sectoral modes are the ridge labeled $\lambda = 0$. The tesseral modes have smaller frequencies in magnitude, with each ridge moving closer to zero frequency as the latitudinal order $\lambda$ increases. The first set of tesseral modes with $\lambda = 1$ are labeled. As expected, all of the equatorial Rossby waves have a negative frequency and thus propagate in the retrograde direction. Ridges with $\lambda = 2, 3$, and $4$ are also visible. Higher-order ridges are less clear because the spacing between adjacent ridges falls below their line widths, leading to blending of the individual peaks. This blurring is more obvious in Figure \ref{fig:radispectra}b, which displays a cut through the radial-vorticity power spectra at $m = 2$ for $0.595 \, R_{\odot}$, summed over all $l$. One can clearly identify peaks matching modes for $m=2$ and $l = 2,3,4,5,$ and 6. In addition to these distinct peaks, at low frequencies, there is a collection of modes with $l > 6$ that have blended together. The green dotted line in Figure~\ref{fig:radispectra}b denotes a contribution from a mode with a different $m$-value, specifically the $m = 3$ sectoral mode. This ``leakage" results from an extremely weak misalignment of the mean angular momentum vector with the axis of the spherical coordinate system caused by numerical noise in the simulation. 

Modes with $\lambda$ as high as 10 are visible in the power profiles for individual spherical harmonic components, as shown in Figure \ref{fig:linecuts} for $m = 2$ and $m = 6$. Each panel displays, from left to right, peaks with $\lambda = 0$ to $\lambda = 10$. The separation between the peaks decreases with increasing $\lambda$, but they are all clearly separated from the zero-frequency feature. With this in mind, the radiative interior of our model features far more equatorial Rossby-wave modes than are evident in the latitudinally summed spectral images.   

\linecuts

 Almost all of the horizontal power in the radiative interior is in the inertial band of frequencies, and the power profiles in Figures \ref{fig:radispectra} and \ref{fig:linecuts} rise up to eight orders of magnitude above the power background, indicating that these equatorial Rossby waves are the most prominent contributor to the horizontal velocity components $v_{\theta}$ and $v_{\phi}$. The rms velocity summed across all modes has a typical speed of 50 cm s$^{-1}$, which explains the enhanced velocities in the radiative interior noted by \citet{2015ApJ...813...95L} and \citet{Matilsky2022} and suggests that the radiative interior is not as quiescent as is often believed.

For completeness, we note that all of the temporal spectra for individual spherical harmonic components (see Figure~\ref{fig:linecuts}) display a common power peak at a slightly negative frequency that is very near zero. In the spectra for $m=6$ (panel b) this feature appears around -7 nHz (or $\omega/\Omega_0 \approx 5 \times 10^{-3}$). This power feature is is likely due to the dynamo cycle which has a similar frequency \citep[see][]{Matilsky2022, Matilsky2023}.

\subsection{Latitudinal Eigenfunctions}

\eigenfunctions

An eigenmode for a Rossby wave living on a 2D spherical surface undergoing solid-body rotation has a single spherical harmonic for its eigenfunction and an eigenfrequency given by Equation~(\ref{eq:disp}). While the frequencies of the Rossby waves in our model are described well by this equation, small discursions from spherical symmetry in the radiative interior, such as differential rotation (see the inset in Figure \ref{fig:diff_profile}) and magnetism, cause small perturbations to both the eigenfrequencies and the eigenfunctions. Thus, we anticipate that a given horizontal eigenfunction will be dominated by a single spherical harmonic with a given harmonic degree $l_0$ ($Y_{l_0}^m$), but will also contain weak contributions from spherical harmonics $Y_l^m$ with the same azimuthal order $m$ and nearby values of the harmonic degree $l \approx l_0$. Figure~\ref{fig:linecuts} clearly evinces such behavior. A spectrum for a specific harmonic degree has not only a single large peak, but also a sequence of small subsidiary peaks at the frequencies associated with other harmonic degrees.  The upper panel of Figure~\ref{fig:eigenfunctions} shows the power spectra in radial vorticity for modes with $m = 2$, plotted versus frequency and harmonic degree $l$. Though most power sits in the dominant spherical harmonic $Y_{l_0}^m$, there is some power that spreads across other $l$ values for each mode. The lower panel displays power integrated around a narrow frequency band centered around the appropriate peak for three modes with $m=2$ and $l_0 = 3, 4,$ and 5. It is clear that the power peaks at the harmonic degree $l_0$ of the dominant spherical harmonic component, but the eigenfunctions have small perturbations (probably due to the differential rotation) that spread significant power over harmonic degrees within $\pm2$ of the dominant degree.

%% file: S5-CZWaves.tex
\section{Waves in the Convection Zone} \label{sec:CZWaves}

\czwaves

The right-hand column of Figure \ref{fig:Wave spectra summary} reveals a wide variety of different inertial oscillations present in the convection zone. To begin the exploration of these modes, Figure \ref{fig:czwaves} illustrates the power in radial vorticity at three different radii within this region. The top row is the sum over the power in equatorially symmetric modes with $\lambda = 0$ and $\lambda = 2$, while the bottom row is the sum over the power in the antisymmetric modes $\lambda = 1$ and $\lambda = 3$. Filled black circles mark equatorial Rossby waves (Section \ref{sec:czrw}), numbers 0--2 mark thermal Rossby waves (Section \ref{sec:thermal}), roman numeral I marks the retrograde mixed mode (Section \ref{sec:czrw,n1}), and roman numerals II--VI mark HFR modes and their latitudinal overtones (Section \ref{sec:hfr}). 

\subsection{Equatorial Rossby Waves}
\label{sec:czrw}

Within the convection zone, equatorial Rossby waves are still present, though they are far less prominent than in the radiative interior due to the presence of other convective and wave motions. From the right-hand column of Figure \ref{fig:Wave spectra summary}, we note that the sectoral modes for $2 \leq m \leq 8$ remain, along with several tesseral modes for $\lambda \leq 4$. This figure is averaged over the entire convection zone, so these modes are present somewhere in that region and not necessarily significant throughout.

In Figure \ref{fig:czwaves}, the black dots denote the theoretical values given by Equation~(\ref{eq:disp}). Because we do not know where these modes live, and since the shift appears to be small, we have opted not to apply a Doppler shift. At the base of the convection zone (left-hand panels), sectoral and tesseral modes are present and distinct for roughly $m < 11$. At the radius closest to the surface (right-hand panels), the only modes that are clearly visible are the sectoral modes for roughly $m \leq 5$ (Figure \ref{fig:czwaves}c), $\lambda = 1$ tesseral modes for $m < 7$ (Figure \ref{fig:czwaves}f), and the $\lambda = 2$ tesseral modes for $m<3$ (Figure \ref{fig:czwaves}c). Further discussion of mode properties is deferred to the Discussion (Section \ref{discussion-rework}).

\subsection{Thermal Rossby Waves}
\label{sec:thermal}

\thermallines

Like equatorial Rossby waves, thermal Rossby waves result from the conservation of potential vorticity \citep[e.g.][]{1968RSPTA.263...93R, 1970JFM....44..441B}. Rather than featuring motions confined to a spherical shell (i.e., $v_{\theta}$ and $v_{\phi}$), thermal Rossby waves manifest as prograde-propagating equatorial convective columns aligned with the rotation axis, hence possessing velocity components in all three directions, but with the dynamically active components being in longitude, $v_\phi$, and in the cylindrical radius, $v_r \sin\theta+v_\theta\cos\theta$ \citep{2022ApJ...932...68H, 2023ApJ...958...48J}. These waves can exist wherever the Coriolis force rivals or dominates buoyancy. Hence, they appear in the convection zone of our model and are excluded from the radiative interior. 

The gravest frequency thermal Rossby wave, here called a Busse mode, is the one that lacks latitudinal nodes in $v_\phi$ \citep{1970JFM....44..441B}. Thus, the wave manifests as a parade of convective columns wrapped around the equator, with each column consisting of a single, axially-aligned roll with unidirectional spin. Because of this latitudinal dependence, in radial vorticity each convective column appears anti-symmetric across the equator and contributes to the spherical harmonic components with odd values of $\lambda$. In the lower panels of Figures \ref{fig:Wave spectra summary} and \ref{fig:czwaves}, this prograde-propagating thermal Rossby wave appears as the positive frequency prong (labeled ``0" in Figure \ref{fig:czwaves}) that rises in frequency and asymptotes toward the frame rotation rate $\Omega_0$ as the azimuthal order increases. \citet{2022ApJ...932...68H} calculated the eigenfrequencies for a neutrally stable layer in a local, equatorial, $f$-plane model. We have adapted their calculation for the depth of the convection zone appropriate for the simulation; the resulting eigenfrequencies are overplotted in Figure~\ref{fig:czwaves} as filled black squares. Because the Busse mode lives in the convection zone near the equator, which is rotating faster than the frame rate, we apply a Doppler shift to these eigenfrequencies. We do not know the specifics of the radial eigenfunction, so we cannot easily pick a theoretically motivated $\Delta \Omega$. Instead, we take the volume-averaged equatorial rotation rate (between $\pm 15 ^{\circ}$ latitude) across the entire convection zone, resulting in $\Delta \Omega = 0.015 \, \Omega_0$.

The narrow line widths of Busse's thermal Rossby waves suggest that the modes are stable and largely linear, with only weak nonlinear coupling to other convective modes. Figure \ref{fig:thermallinecuts}a displays select power profiles with $\lambda = 1$ for $m = 2, 4$, and $6$ at a radius of $0.889 \, R_{\odot}$, binned down in frequency by a factor of 32. We fit these binned profiles to a Lorentzian with a linear background term, and these fits are listed in Table \ref{tab:fitting} and overplotted in Figure \ref{fig:thermallinecuts}a with black lines.  Widths ranging between $0.03\,\Omega_0$ and $0.08 \,\Omega_0$ indicate lifetimes of 10--30 rotation periods. The quality factor, $Q$---i.e., the ratio of the line width to the mode frequency---is about ten across all eight modes. Further details regarding line fitting are found in Appendix \ref{sec:linefitting}.

In addition to Busse modes, i.e., the fundamental thermal Rossby wave consisting of a single, axially aligned roll in latitude (with a radial vorticity that is antisymmetric across the equator), there is a clear signature in our spectra for the first latitudinal overtone. Such thermal Rossby waves consist of two counter-rotating rolls stacked in latitude, with one node at the equator in $v_{\phi}$. Since thermal Rossby waves of this type were first explored by \citet{1968RSPTA.263...93R}, we refer to these modes as Roberts modes, and they are labeled ``1" in Figure \ref{fig:czwaves}. Such modes have a radial vorticity that is symmetric across the equator and thus appear in the even values of $\lambda$. We have overplotted the eigenfrequency calculations of \citet{2022AA...662A..16B} with open squares, with an applied Doppler shift of $0.015 \, \Omega_0$. \citet{1968RSPTA.263...93R} first examined these modes in a Boussinesq system, but \citet{2023ApJ...958...48J} have recently considered analytical solutions in a stratified atmosphere. Such modes have also been detected by \citet{2022AA...662A..16B} in their linear-mode calculations where they have noted that this family of waves smoothly transitions from the branch of prograde thermal Rossby waves to a branch of retrograde inertial waves with one radial node in radial vorticity as the azimuthal order $m$ passes through zero from positive to negative values. We find the same (Section \ref{sec:czrw,n1}) and, following \citet{2022AA...662A..16B}, refer to this behavior where a prograde branch and a retrograde branch are connected as mixed modes.

Figure \ref{fig:thermallinecuts}b shows select line profiles for the Roberts modes. The modes appear in both $\lambda = 0$ and $\lambda = 2$ and appear clearest in $\lambda = 2$ at a radius near the surface of $0.935 R_{\odot}$. Because the line profiles are asymmetric about the peak frequency and the background power has a strong nonlinear frequency dependence, we have elected not to fit them for this work.  

Beyond these two easily identifiable types of thermal Rossby waves, we do find evidence of further latitudinal overtones, both symmetric and antisymmetric across the equator. In Figure \ref{fig:czwaves}f, there is a prograde, higher frequency ridge of power (labeled ``2") that slopes downwards and asymptotes towards the rotation rate. This is likely a thermal Rossby wave with two latitudinal nodes in $v_\phi$, or three stacked columns, that is antisymmetric in radial vorticity (i.e., three latitudinal nodes in radial vorticity). In Figure  \ref{fig:thermallinecuts}b, an additional symmetric mode appears as a high-frequency bump in power in $m = 6$ (near a frequency of $\omega = 1.3 \, \Omega_0$), labeled ``3" here and in Figure \ref{fig:czwaves}. The eigenfunctions of these columnar waves are not spherical harmonics, so we expect latitudinal overtones to be spread across multiple spherical-harmonic components. Because this high-frequency bump appears in an equatorially symmetric spherical harmonic, this feature could indicate the presence of the mode with four columns stacked in latitude and three latitudinal nodes in $v_{\phi}$. (In radial vorticity, there would be four latitudinal nodes.) In Section~\ref{sec:hfr_disc} we speculate that these high-order modes correspond to the prograde branches of mixed HFR waves.

\subsection{Retrograde Mixed Modes}
\label{sec:czrw,n1}

Throughout the tachocline and convection zone, there exists in the spectra for equatorially symmetric vorticity a strong ridge of power at low $m$ with frequencies that lie between the ridges for the $\lambda = 0$ and $\lambda = 1$ equatorial Rossby modes (top rows of Figures \ref{fig:Wave spectra summary} and \ref{fig:czwaves}, labeled with roman numeral I). \citet{2022AA...662A..16B} referred to these modes as sectoral equatorial Rossby modes with one radial node ($n = 1$) and describe them as having a dominant $\lambda = 0$ component in $v_{\theta}$ and a dominant $\lambda = 1$ component in $v_{\phi}$, which would result in the equatorially symmetric signature that we see in the radial vorticity. The Doppler-shifted values ($\Delta \Omega = 0.015 \Omega_0$) from Table 2 of \citet{2022AA...662A..16B} are overplotted as empty black circles and match the ridges rather well.

Figure \ref{fig:thermallinecuts}c displays line profiles for the $m = 2, 4,$ and $6$ modes for $\lambda = 0$ at a radius of $0.935 \, R_{\odot}$. The equatorial Rossby wave $(l, m) = (2, 2)$ is visible as the labeled tiny peak on the far left, emphasizing that the mixed modes are significantly wider and larger in amplitude. We again perform a Lorentzian fit with a linear background. These fits are overplotted as solid black lines in Figure \ref{fig:thermallinecuts}c, and the values are listed in Table \ref{tab:mixedfitting}. The widths of these peaks are typically $0.05 \, \Omega_0$, which corresponds to mode lifetimes of 20 rotations, and a $Q$-factor of between 2 and 10. While these fits provide a good fit to the core of the line profile, and hence a reasonable estimate for the linewidth, the fits to the wings and the power background are clearly inadequate. In particular, the wings are slightly asymmetric. Since we do not currently have a good physical argument for this asymmetry, we have elected to fit a symmetric Lorentzian with a linear background rather than fit an asymmetric profile. 

As discussed in \citet{2022AA...662A..16B}, these modes (in this work, corresponding to Ridge I) are mixed modes. Specifically these retrograde waves are smoothly joined through $m = 0$ to the Roberts modes (Ridge 1 in the upper panels of Figure \ref{fig:czwaves}), which have one node in latitude at the equator in the variable $v_\phi$ (Section \ref{sec:thermal}). In our spectra, where the prograde waves are presented with positive frequencies and the retrograde waves with negative ones, the mixed nature appears as a common absolute frequency $\abs{\omega}$ for the $m=0$ mode of each branch. The $(l, m) = (2, 0)$ mode is plotted in Figure \ref{fig:thermallinecuts}c, and it occurs at the same absolute frequency as the corresponding $m = 0$ mode in Figure \ref{fig:thermallinecuts}b.

\subsection{HFR Modes}
\label{sec:hfr}

\hfrlinecuts
\hfrpoloidal

Following the observational discovery of HFR modes on the Sun by \citet{2022NatAs...6..708H}, \cite{2022ApJ...934L...4T} provided numerical evidence to support the identification of HFR waves as a separate class from equatorial Rossby waves. Using a linear eigensolver for the Boussinesq equations, \citet{2022ApJ...934L...4T} found a class of equatorially antisymmetric modes with similar frequencies to those in \citet{2022NatAs...6..708H} and featuring larger poloidal (i.e., vertical) kinetic energy than Rossby waves. \citet{2023ApJS..264...21B} used an anelastic solver and generated qualitatively similar high-frequency modes, along with an equatorially symmetric branch that has not yet been observed. \citet{2023arXiv231110414B} corroborated the previous eigensolver results for the primary set of anti-symmetric modes.

We find evidence for both the symmetric and antisymmetric branches of the HFR modes. In Figure \ref{fig:czwaves}, there are three anti-symmetric bands (labeled II, IV, and VI) and two symmetric bands (III, V) of retrograde-propagating inertial waves occurring at frequencies of higher magnitude than the equatorial Rossby waves. The HFR waves observed by \citet{2022NatAs...6..708H}, marked as plus signs on the figure with a Doppler shift of $\Delta \Omega = 0.015~\Omega_0$, are anti-symmetric in radial vorticity and found near the solar surface. The underlying anti-symmetric band of power in our spectra is most likely these HFR modes. However, since the nature of the HFR modes is still unclear from both an observational and theoretical standpoint, this identification is based purely on the equatorial symmetry and the similarity in frequencies.

In addition to this anti-symmetric ridge, we see a strong ridge of power (III) in the spectra for modes that are symmetric across the equator, for $\lambda = 0$ and $\lambda = 2$, that is similar to the symmetric $\lambda = 0$ ridge seen by \citet{2023ApJS..264...21B}. Beyond these two prominent ridges, we note additional bands of high-frequency power (IV, V, VI) that to our knowledge have not been reported previously. These features are also visible in Figure \ref{fig:hfrlinecuts}, which displays the corresponding power profiles in radial vorticity for $m = 9$ and $\lambda = 0$ through $4$. These additional enhancements in power are slight and extremely broad; hence, if modal, have an extremely low $Q$ factor. 

Figure \ref{fig:hfrpoloidal} displays power spectra for the poloidal velocity $|v_\theta|^2 + |v_r|^2$, summed over two different radii, with the five HFR ridges again marked by roman numerals II-VI and I denoting the retrograde mixed mode. Because the equatorial symmetries of $v_{\theta}$ and $v_r$ are opposite, the modes are denoted by the symmetry consistent with the radial vorticity; the ``symmetric modes" are the sum of $|v_{\theta}|^2$ over $\lambda = 0, 2$ and $|v_r|^2$ over $\lambda = 1, 3$, and the antisymmetric modes are the converse. Values from \citet{2022AA...662A..16B} are marked with open circles, and equatorial Rossby waves are marked with closed circles. Panel (a) shows power with equatorial symmetry, with values from \citet{2023ApJS..264...21B} overplotted as stars, while Panel (c) shows the antisymmetric power, with observations from \citet{2022NatAs...6..708H} overplotted as plus signs, along with numerical values from \citet{2022ApJ...934L...4T} (triangles) and \citet{2023ApJS..264...21B} (times signs), again all Doppler shifted by $0.015 \, \Omega_0$. We note that Ridges II and III have the same equatorial symmetry and similar frequencies as previous observations and numerical calculations, with particular concordance with the results of \citet{2023ApJS..264...21B}.

A key feature of all of these modes is a significant vertical velocity component, both deep within the convection zone and at the surface. This property has been recognized in previous studies \citep{2022ApJ...934L...4T,2023arXiv231110414B} and is quite evident from the spectra in Figure \ref{fig:hfrpoloidal}. Panels (b) and (d) of Figure \ref{fig:hfrpoloidal} show $m = 9$ power profiles for $v_{\theta}$ and $v_r$ spectra at two different radii. The $v_{\theta}$ spectra have been summed over equatorially symmetric (antisymmetric) power $\lambda = 0, 2$ ($\lambda = 1, 3$). $v_r$ exhibits the opposite symmetry, so the symmetric modes are the sum of $\lambda = 1, 3$ while the anti-symmetric modes are the sum of $\lambda = 0, 2$. We note that while $v_{\theta}$ is a couple of orders of magnitude greater than $v_r$ at the surface, the difference reduces to less than one order of magnitude at the deeper radial slice, which qualitatively matches previous findings \citep{2022ApJ...934L...4T,2023arXiv231110414B} 

It is highly probable that these five ridges (II--VI) and the mixed retrograde wave (I) are all in the same family, with higher-frequency ridges corresponding to modes of higher latitudinal order. Further analysis is deferred to Section \ref{sec:hfr_disc}.

\subsection{Impact of Differential Rotation on Convection}

\diffspectra

The final feature that we identify in the spectra is the convection, which exhibits a Doppler shift due to the radial variation of the mean rotation rate. In Figure \ref{fig:Wave spectra summary}, this feature has been averaged over several different radii and manifests as the smear of power around and above zero frequency. Figure \ref{fig:diffspectra} shows the spectra in radial vorticity, for $\lambda = 0$, out to $m = 80$ at three different radii in the convection zone. The overplotted dashed black line shows the expected frequency resulting from the Doppler shift due to the difference between the local mean rotation rate and the frame rate $\Omega_0$. At the base of the convection zone, the equatorial rotation rate is almost the same as the frame rate, so the convective power remains near zero frequency for all $m$. Higher in the convection zone, the equatorial rotation rate rises above the frame rate by as much as $0.036 \, \Omega_0$, resulting in a significant Doppler-shift and the sloped distribution of power present in the center and right panels of Figure \ref{fig:diffspectra}. The keen-eyed reader will note the sectoral equatorial Rossby waves and the retrograde mixed mode in the lower left of each panel, along with a squished thermal Rossby-wave branch (prograde frequencies) and an equatorially symmetric HFR branch (retrograde frequencies) in Panel (c). 

%% file: S6-Discussion.tex
\section{Discussion} \label{sec:discussion}

\summarytable

This simulation simultaneously features a wide variety of self-consistently excited inertial oscillations, some of which have been seen previously in observations and models, and some of which have not. Table \ref{tab:summarytable} lists the waves discussed in this work, where they exist in the simulation, if they have been observed, and whether they have been seen in recent models. Equatorial Rossby waves in the radiative interior, tesseral equatorial Rossby waves in the convection zone, and high-frequency retrograde vorticity waves have not been previously noted in a 3D convection simulation such as this one. Because this model has a solar-like rotation profile (fast equator and slow poles) complete with stable and unstable regions and a tachocline, it provides a useful tool to investigate how these waves manifest and where their wave cavities reside. 

\subsection{Equatorial Rossby Waves} \label{discussion-rework}

\radialres

\subsubsection{Are There Two Wave Cavities?}

\freqshift

In their study of the radial behavior of r-modes, \citet{1986SoPh..105....1W} numerically calculated the first-order frequency correction to Equation~(\ref{eq:disp}) for a solar interior model, with a stably stratified radiative interior beneath an unstable convection zone. They found that there are two separate wave cavities, one in the radiative interior and another in the convection zone, with the frequency correction being of opposite sign in the two regions. Specifically, if we add the frequency correction that arises from stratification, $\delta\omega_{\rm strat}$ to Equation~(\ref{eq:disp}),
\begin{equation}
    \omega = m\Delta\Omega - \frac{2m\Omega_0}{l(l+1)} + \delta\omega_{\rm strat} \;,
\end{equation}

\noindent \citet{1986SoPh..105....1W} found that the the frequency correction is positive for the convection zone cavity, $\delta\omega_{\rm strat}>0$, and negative for the cavity in the stably stratified radiative interior, $\delta\omega_{\rm strat}<0$. Further, \citet{1986SoPh..105....1W} found that such cavities can support radial overtones with differing numbers of nodes in radius. In the following subsections we explore both of these possibilities: that the Rossby modes that we have observed in our numerical simulation live in two distinct cavities and whether radial overtones might be present.

A resonant normal mode has a single frequency that is independent of spatial position. Hence, we can attempt to distinguish between modes in the radiative interior and modes in the convection zone by their frequencies. In addition to the frequency shift mentioned previously that arises from the nature of the stratification (negative in the radiative zone and positive in the convection zone), we expect a Doppler shift based on the difference in the rotation rate between the two zones. Because the radiative interior is rotating slightly slower than the frame rate of our model, we expect a negative $\Delta \Omega$ correction in Equation~(\ref{eq:disp}). Conversely, the convection zone is rotating faster than the frame rate, necessitating a positive $\Delta \Omega$ correction. The expected Doppler shifts thus have the same sign as the frequency shift identified by \citet{1986SoPh..105....1W}.

Figure \ref{fig:radialres} displays power profiles for the $(l,m) = (4,4)$ sectoral mode at full frequency resolution at the base of the radiative interior and in the lower convection zone. The purple dashed line marks the expected frequency for modes in a region rotating at the frame rate (where $\Delta \Omega = 0$) and with no stratification correction, $\delta\omega_{\rm strat} = 0$. We can clearly see that the power profile within the radiative interior has a mean frequency that is higher in magnitude than the purple dashed line (more negative), while the convection-zone profile has its mean at a lower-magnitude frequency (less negative).

With this expected frequency difference in mind, we average the power in each mode over the radiative interior and over the convection zone and calculate the first frequency moment $\langle \omega \rangle$ of the power distribution separately in each region (the angular brackets here referring to a power-spectrum-weighted average). Figure \ref{fig:freqshift} shows the difference, $\langle \Delta \omega \rangle = \langle \omega \rangle_{cz} - \langle \omega \rangle_{ri}$, in the mean frequencies for the convection zone and radiative interior, scaled by the 2D dispersion relation value $\omega_{\rm 2D} = 2 m \Omega_0/l(l+1)$. Because $\langle \Delta\omega \rangle$ is always positive, the convection zone modes always have a lower frequency (less negative) than those in the radiative interior, as predicted by \citet{1986SoPh..105....1W}. There is not a clear relationship between frequency difference and latitudinal node number $\lambda$, but for the sectoral modes, $\langle \Delta \omega \rangle$ does increase with $m$.

The distinct shift in frequencies between equatorial Rossby waves in the radiative interior and those in the convection zone suggests that the equatorial Rossby waves are behaving as \cite{1986SoPh..105....1W} predicted. The waves are split into two different families. One family has frequencies that are more negative than the classic dispersion relation would indicate, and these live in the radiative interior, i.e., they propagate in the radiative interior and are evanescent in the convection zone. Conversely, the other family has frequencies that are less negative than the classic 2D dispersion relation and live in the convection zone (i.e., they propagate in the convection zone and are evanescent in the radiative interior). The frequency shift between these two families is tiny (on the order of $0.01 \, \Omega_0$) and would likely have gone unnoticed in observations. The observed linewidths are typically 20--40 nHz \citep[e.g.,][]{2018NatAs...2..568L}, which corresponds to 0.3--0.6 $\Omega_0$; hence, an observed spectral peak is too broad to separate the two potential peaks. Furthermore, since the deeply seated modes that live in the radiative interior are likely to have lower amplitude at the solar surface due to their evanescence in the convection zone, current photospheric observations are inherently less sensitive to them. 

\subsubsection{Do We See Radial Overtones?}

In addition to splitting the power into two spectral peaks due to potentially distinct stable-zone and convection-zone mode cavities, the power profiles for the equatorial Rossby waves are clearly not Lorentzian and may actually be the superposition of many under-resolved peaks. The spectrum for the $(l,m) = (4,4)$ mode as seen in the radiative interior (Figure \ref{fig:radialres}) evinces many narrow peaks. To our eyes, these do not look like realization noise spikes as many of the peaks smoothly span multiple nearby frequencies. However, we admit that the stochastic excitation of these modes has only begun to be explored \citep[see][]{Philidet2023}, and frequency correlations in the realization noise are poorly understood.

If real (and not noise), the closely spaced, under-resolved peaks could be radial overtones of the equatorial Rossby waves in the radiative interior. Because they exist in a highly stratified, stable environment, these radial overtones would be very closely packed in frequency with a spacing likely to be inversely proportional to the square of the buoyancy frequency \citep[see][]{2017aofd.book.....V,2023A&A...671A..91A}. Our non-dimensional frequency resolution is $1.1 \cross 10^{-4}$, but \citet{1986SoPh..105....1W} estimate the radial splitting between peaks to be as small as $2 \cross 10^{-4}$. We do not have the frequency resolution, nor with the present modelling results can we average over sufficient realizations for noise reduction, to make a definitive statement or to distinguish between fine-scale modal structure and realization noise.

The issue of whether radial overtones are present is further complicated by the presence of differential rotation.  It is well-known that shear flows permit the existence of additional families of inertial waves \citep[e.g.,][]{Kuo1949, Mack1976}, such as the critical-latitude modes \citep{2021AA...652L...6G, 2020AA...642A.178G}. These additional modes also are expected to have a dense spectra with peaks that blend together \citep{Philidet2023} and without detailed information about the eigenfunctions may be very difficult to disentangle from radial overtones.

\subsection{HFR Modes May Be the Retrograde Branch of Thermal Rossby Waves} \label{sec:hfr_disc}

\mixedmodes

In Section \ref{sec:hfr}, amongst the retrograde propagating waves, we find two equatorially symmetric bands of power and three anti-symmetric bands of power (labelled II--VI in Figures~\ref{fig:czwaves} and \ref{fig:hfrpoloidal}). The lowest-frequency symmetric and anti-symmetric bands are consistent with previous HFR mode observations and simulations, but the additional ridges are new. These lesser features are likely latitudinal overtones of these HFR modes. Given that we are seeing the same features across different latitudinal orders (Figure \ref{fig:hfrlinecuts}), the eigenfunctions of these waves are linear combinations of the spherical harmonics.

While we suspect that these five ridges are all HFR modes, there are still open questions about the physical mechanism that produces them. \citet{2023ApJ...958...48J} used an analytic model to study the radial and latitudinal propagation of thermal Rossby waves in an isentropically stratified atmosphere. For modes with latitudinal propagation, they find a new set of retrograde-propagating inertial waves whose eigenfrequencies and latitudinal behavior bear a qualitative resemblance to the HFR modes---see Figure 5 in \citet{2023ApJ...958...48J}. These retrograde inertial waves are the retrograde branch of the prograde thermal Rossby waves. They only exist for waves that possess latitudinal nodes, and for $m=0$ the motions consist of rolls with axes aligned with lines of constant latitude. For $m\neq0$, the modes have fully 3D motions. Notably, these modes have similar latitudinal behavior to the potential HFR modes that we have identified in our simulation, with latitudinal overtones occurring at higher and higher (negative) frequencies without limit.

If the HFR modes are the same type of mode as identified by \citet{2023ApJ...958...48J}, then the HFR modes are in the thermal Rossby-wave family and are mixed modes. By this we mean that the prograde thermal Rossby-wave branch smoothly transitions to a retrograde HFR mode as the azimuthal order $m$ passes through zero from positive to negative values. 

Figure \ref{fig:mixedmodes} plots the power spectra for the poloidal velocity power, summed over $\lambda = 0$--3, averaged across two radial slices near the surface. Prograde-propagating waves have been shown with negative values of both the azimuthal order $m$ and frequency $\omega/\Omega_0$ to better make mixed-mode relationships clear. The prograde-propagating thermal Rossby waves are again labeled 0, 1, and 2, while the retrograde branches are labeled with roman numerals. For example, the previously discussed mixed mode (Section \ref{sec:czrw,n1}) is denoted with open circles for the retrograde modes (labeled I) and open squares for the prograde modes, i.e., Roberts modes (labeled 1), with numerical values obtained from \citet{2022AA...666A.135B}. The Busse mode (labeled 0) is again marked by closed squares with numerical values from \citet{2022ApJ...932...68H}.

In Figure \ref{fig:mixedmodes}, we note that the anti-symmetric HFR branch (II) seems to flow into the prograde thermal Rossby wave with two latitudinal nodes (2), which directly relates the primary HFR observations to the previously noted latitudinal overtone. Additionally, while it is not as visible in this figure, we remind the reader that we identified a potential thermal Rossby wave with high frequency and three latitudinal nodes in Figure \ref{fig:thermallinecuts}b (see Section \ref{sec:thermal}). This overtone was most visible for high-$m$ values, and has a frequency of opposite sign but suspiciously similar magnitude to numerical HFR-mode results. Ridge III in Figure \ref{fig:mixedmodes} could potentially transition into this higher-order thermal Rossby wave. Ridges IV--VI may exhibit similar patterns, though it is difficult to make a definitive conclusion due to their low amplitudes.

Furthermore, the retrograde inertial wave identified by \citet{2022AA...662A..16B} has already been shown to be a mixed mode that transitions (when prograde propagating) to a thermal Rossby wave, specifically to one with one node in $v_{\phi}$  in latitude. The equatorial symmetry of this mixed mode and its frequency suggests that it might be a member of the HFR family. If true, the picture becomes simple.  There are only two unique families of inertial waves in the convection zone. We have equatorial Rossby waves which are primarily horizontal in motion. Separately there exists a sequence of latitudinal overtones of a truly 3D mode that is a prograde thermal Rossby wave ``mixed" with a retrograde HFR mode, with the Busse mode as the gravest member. The mode found by \citet{2022AA...662A..16B}, is just one member of this series. 

\subsection{Absent Inertial Waves}

While this simulation features many types of inertial waves, several varieties are notably missing. We find no evidence of critical-latitude modes \citep[e.g.,][]{2020AA...642A.178G,2021AA...652L...6G}, most likely due to the relatively weak differential rotation in this model. Additionally, we find no evidence of high-latitude modes \citep[e.g.,][]{2021AA...652L...6G}, also potentially due to the weak differential rotation. Through thermal wind balance, weak differential rotation leads to weak latitudinal entropy gradients, and the high-latitude modes have been demonstrated to reach large amplitude due to baroclinic instability \citep{2022AA...662A..16B}. However, we further acknowledge that the spectral decomposition into spherical harmonics that we perform is not the most robust way to identify high-latitude features. Finally, as discussed in Sections \ref{sec:methods} and \ref{sec:RIWaves}, there is no evidence of the splitting of Rossby waves into fast and slow MHD Rossby waves by the magnetic field. This absence is most certainly due to the rather weak magnetic fields generated by this particular dynamo, and the resulting magnetic splittings are too small to discriminate (see Figure~\ref{fig:simparams}c).

%% file: S7-Conclusion.tex
\section{Conclusion}

In this work, we presented a 3D numerical simulation of the upper radiative interior and convection zone that features a wide variety of inertial oscillations. We summarize our primary findings as follows:

\textbf{1. Rossby waves in the radiative interior:} 
We presented a rich wave-field of equatorial Rossby waves in the radiative interior and demonstrated that they account for most of the horizontal motion in that region. Their presence raises questions about their impact on dynamo and transport properties in a region of the Sun that has historically been considered rather quiescent. 

\textbf{2. Rossby-wave cavities:} We found both sectoral and tesseral equatorial Rossby waves throughout the convection zone that seem to occur at frequencies distinct from those in the radiative interior. This implies that there are potentially two unique families of equatorial Rossby waves living in separate radiative-interior and convection-zone cavities. The presence of both sectoral and tesseral equatorial Rossby waves of varying node number throughout the domain bodes well for their potential helioseismic utility, but it does raise some questions. We do not yet know how these modes are excited and whether this mechanism differs between the radiative interior and convection zone. We also need to develop a better understanding of the conditions under which these waves live and how they interact with each other. Stratification and superadiabaticity in particular may play a significant role in controlling the mode frequencies and determining the amplitude of any given radial overtones.

\textbf{3. HFR modes and thermal Rossby waves:} We noted the presence of both previously observed and modeled branches of HFR modes along with three additional branches that are probably their latitudinal overtones. These modes seem to be mixed with thermal Rossby waves. If true, the theoretical picture becomes simple and unified. The Busse modes, with no nodes in latitude, are the fundamental mode. The previously noted mixed mode, which relates Roberts modes to a 3D retrograde oscillation, is the first latitudinal overtone. The anti-symmetric HFR modes, mixed with prograde thermal Rossby waves with two nodes in latitude, are the second latitudinal overtone, and so on. 

%% file: Z1-Linefitting.tex
\appendix

\section{Line Fitting} \label{sec:linefitting}

Radial-vorticity spectral line profiles were fit for a selection of thermal Rossby waves (ridge 0) and the retrograde mixed mode (ridge I). Power profiles were fit with the sum of a Lorentzian and a linear background term:

\begin{equation} \label{eq:lorentzian}
    f(\omega | A, \omega_0, \gamma, B_0, \alpha) = \frac{A}{1 + \left(\frac{\omega - \omega_0}{\gamma}\right)^2} + B_0 + \alpha \omega,
\end{equation}

\noindent where $A$ is the amplitude, $\omega_0$ is the central frequency, $\gamma$ is the width, and $B_0$ and $\alpha$ are linear background parameters. We used the least squares curve\_fit routine from the Python SciPy package. The results of this fitting are given in Tables \ref{tab:fitting} and \ref{tab:mixedfitting}. The central frequency and width parameters have been non-dimensionalized by the rotation rate for ease of comparison with other studies.

\fitting
\mixedfitting

\newpage

%% file: main.bbl
\begin{thebibliography}{}
\expandafter\ifx\csname natexlab\endcsname\relax\def\natexlab#1{#1}\fi
\providecommand{\url}[1]{\href{#1}{#1}}
\providecommand{\dodoi}[1]{doi:~\href{http://doi.org/#1}{\nolinkurl{#1}}}
\providecommand{\doeprint}[1]{\href{http://ascl.net/#1}{\nolinkurl{http://ascl.net/#1}}}
\providecommand{\doarXiv}[1]{\href{https://arxiv.org/abs/#1}{\nolinkurl{https://arxiv.org/abs/#1}}}

\bibitem[{{Albekioni} {et~al.}(2023){Albekioni}, {Zaqarashvili}, \&
  {Kukhianidze}}]{2023A&A...671A..91A}
{Albekioni}, M., {Zaqarashvili}, T.~V., \& {Kukhianidze}, V. 2023, \aap, 671,
  A91, \dodoi{10.1051/0004-6361/202243985}

\bibitem[{{Alshehhi} {et~al.}(2019){Alshehhi}, {Hanson}, {Gizon}, \&
  {Hanasoge}}]{Alshehhi2019}
{Alshehhi}, R., {Hanson}, C.~S., {Gizon}, L., \& {Hanasoge}, S. 2019, \aap,
  622, A124, \dodoi{10.1051/0004-6361/201834237}

\bibitem[{{Alvan} {et~al.}(2015){Alvan}, {Strugarek}, {Brun}, {Mathis}, \&
  {Garcia}}]{2015A&A...581A.112A}
{Alvan}, L., {Strugarek}, A., {Brun}, A.~S., {Mathis}, S., \& {Garcia}, R.~A.
  2015, \aap, 581, A112, \dodoi{10.1051/0004-6361/201526250}

\bibitem[{{Aurnou} {et~al.}(2020){Aurnou}, {Horn}, \&
  {Julien}}]{2020PhRvR...2d3115A}
{Aurnou}, J.~M., {Horn}, S., \& {Julien}, K. 2020, Physical Review Research, 2,
  043115, \dodoi{10.1103/PhysRevResearch.2.043115}

\bibitem[{{Azouni} {et~al.}(1985){Azouni}, {Bolton}, \& {Busse}}]{Azouni1985}
{Azouni}, M.~A., {Bolton}, E.~W., \& {Busse}, F.~H. 1985, Geophysical and
  Astrophysical Fluid Dynamics, 34, 301, \dodoi{10.1080/03091928508245448}

\bibitem[{{Bekki}(2023)}]{2023arXiv231110414B}
{Bekki}, Y. 2023, arXiv e-prints, arXiv:2311.10414,
  \dodoi{10.48550/arXiv.2311.10414}

\bibitem[{{Bekki} {et~al.}(2022{\natexlab{a}}){Bekki}, {Cameron}, \&
  {Gizon}}]{2022AA...666A.135B}
{Bekki}, Y., {Cameron}, R.~H., \& {Gizon}, L. 2022{\natexlab{a}}, \aap, 666,
  A135, \dodoi{10.1051/0004-6361/202244150}

\bibitem[{{Bekki} {et~al.}(2022{\natexlab{b}}){Bekki}, {Cameron}, \&
  {Gizon}}]{2022AA...662A..16B}
---. 2022{\natexlab{b}}, \aap, 662, A16, \dodoi{10.1051/0004-6361/202243164}

\bibitem[{{Bhattacharya} \& {Hanasoge}(2023)}]{2023ApJS..264...21B}
{Bhattacharya}, J., \& {Hanasoge}, S.~M. 2023, \apjs, 264, 21,
  \dodoi{10.3847/1538-4365/aca09a}

\bibitem[{Braginsky \& Roberts(1995)}]{BR1995}
Braginsky, S.~I., \& Roberts, P.~H. 1995, Geophysical \& Astrophysical Fluid
  Dynamics, 79, 1, \dodoi{10.1080/03091929508228992}

\bibitem[{{Bretherton} \& {Garrett}(1968)}]{1968RSPSA.302..529B}
{Bretherton}, F.~P., \& {Garrett}, C.~J.~R. 1968, Proceedings of the Royal
  Society of London Series A, 302, 529, \dodoi{10.1098/rspa.1968.0034}

\bibitem[{{Busse}(1970)}]{1970JFM....44..441B}
{Busse}, F.~H. 1970, Journal of Fluid Mechanics, 44, 441,
  \dodoi{10.1017/S0022112070001921}

\bibitem[{{Busse} \& {Hood}(1982)}]{Busse1982}
{Busse}, F.~H., \& {Hood}, L.~L. 1982, Geophysical and Astrophysical Fluid
  Dynamics, 21, 59, \dodoi{10.1080/03091928208209005}

\bibitem[{{Camisassa} \& {Featherstone}(2022)}]{2022ApJ...938...65C}
{Camisassa}, M.~E., \& {Featherstone}, N.~A. 2022, The Astrophysical Journal,
  938, 65, \dodoi{10.3847/1538-4357/ac879f}

\bibitem[{{Chamberlain} \& {Carrigan}(1986)}]{Chamberlain1986}
{Chamberlain}, J.~A., \& {Carrigan}, C.~R. 1986, Geophysical and Astrophysical
  Fluid Dynamics, 35, 303, \dodoi{10.1080/03091928608245897}

\bibitem[{{Clune} {et~al.}(1999){Clune}, {Elliott}, {Miesch}, {Toomre}, \&
  {Glatzmaier}}]{Clune1999}
{Clune}, T., {Elliott}, J., {Miesch}, M., {Toomre}, J., \& {Glatzmaier}, G.
  1999, Parallel Computing, 25, 361, \dodoi{10.1016/S0167-8191(99)00009-5}

\bibitem[{{Cordero} \& {Busse}(1992)}]{Cordero1992}
{Cordero}, S., \& {Busse}, F.~H. 1992, \grl, 19, 733, \dodoi{10.1029/92GL00574}

\bibitem[{{Featherstone} {et~al.}(2021){Featherstone}, {Edelmann},
  {Gassmoeller}, {Matilsky}, {Orvedahl}, \& {Wilson}}]{2021zndo...5774039F}
{Featherstone}, N.~A., {Edelmann}, P. V.~F., {Gassmoeller}, R., {et~al.} 2021,
  {geodynamics/Rayleigh: Rayleigh Version 1.0.1}, 1.0.1, Zenodo,  Zenodo,
  \dodoi{10.5281/zenodo.5774039}

\bibitem[{{Featherstone} \&
  {Hindman}(2016{\natexlab{a}})}]{2016ApJ...818...32F}
{Featherstone}, N.~A., \& {Hindman}, B.~W. 2016{\natexlab{a}}, \apj, 818, 32,
  \dodoi{10.3847/0004-637X/818/1/32}

\bibitem[{{Featherstone} \&
  {Hindman}(2016{\natexlab{b}})}]{2016ApJ...830L..15F}
---. 2016{\natexlab{b}}, \apjl, 830, L15, \dodoi{10.3847/2041-8205/830/1/L15}

\bibitem[{{Fournier} {et~al.}(2022){Fournier}, {Gizon}, \&
  {Hyest}}]{2022AA...664A...6F}
{Fournier}, D., {Gizon}, L., \& {Hyest}, L. 2022, \aap, 664, A6,
  \dodoi{10.1051/0004-6361/202243473}

\bibitem[{{Gill}(1982)}]{Gill1982}
{Gill}, A.~E. 1982, {Atmosphere-Ocean Dynamics} ({Academic Press})

\bibitem[{{Gilman}(1987)}]{Gilman1987}
{Gilman}, P.~A. 1987, \apj, 318, 904, \dodoi{10.1086/165422}

\bibitem[{{Gilman} \& {Glatzmaier}(1981)}]{1981ApJS...45..335G}
{Gilman}, P.~A., \& {Glatzmaier}, G.~A. 1981, \apjs, 45, 335,
  \dodoi{10.1086/190714}

\bibitem[{{Gizon} {et~al.}(2020){Gizon}, {Fournier}, \&
  {Albekioni}}]{2020AA...642A.178G}
{Gizon}, L., {Fournier}, D., \& {Albekioni}, M. 2020, \aap, 642, A178,
  \dodoi{10.1051/0004-6361/202038525}

\bibitem[{{Gizon} {et~al.}(2021){Gizon}, {Cameron}, {Bekki}, {Birch}, {Bogart},
  {Brun}, {Damiani}, {Fournier}, {Hyest}, {Jain}, {Lekshmi}, {Liang}, \&
  {Proxauf}}]{2021AA...652L...6G}
{Gizon}, L., {Cameron}, R.~H., {Bekki}, Y., {et~al.} 2021, \aap, 652, L6,
  \dodoi{10.1051/0004-6361/202141462}

\bibitem[{{Glatzmaier}(1984)}]{1984JCoPh..55..461G}
{Glatzmaier}, G.~A. 1984, Journal of Computational Physics, 55, 461,
  \dodoi{10.1016/0021-9991(84)90033-0}

\bibitem[{Gough(1969)}]{TheAnelasticApproximationforThermalConvection}
Gough, D.~O. 1969, Journal of Atmospheric Sciences, 26, 448 ,
  \dodoi{10.1175/1520-0469(1969)026<0448:TAAFTC>2.0.CO;2}

\bibitem[{{Hanasoge} \& {Mandal}(2019)}]{Hanasoge2019}
{Hanasoge}, S., \& {Mandal}, K. 2019, \apjl, 871, L32,
  \dodoi{10.3847/2041-8213/aaff60}

\bibitem[{{Hanson} {et~al.}(2020){Hanson}, {Gizon}, \& {Liang}}]{Hanson2020}
{Hanson}, C.~S., {Gizon}, L., \& {Liang}, Z.-C. 2020, \aap, 635, A109,
  \dodoi{10.1051/0004-6361/201937321}

\bibitem[{{Hanson} {et~al.}(2022){Hanson}, {Hanasoge}, \&
  {Sreenivasan}}]{2022NatAs...6..708H}
{Hanson}, C.~S., {Hanasoge}, S., \& {Sreenivasan}, K.~R. 2022, Nature
  Astronomy, 6, 708, \dodoi{10.1038/s41550-022-01632-z}

\bibitem[{{Hathaway} \& {Upton}(2021)}]{Hathaway2021}
{Hathaway}, D.~H., \& {Upton}, L.~A. 2021, \apj, 908, 160,
  \dodoi{10.3847/1538-4357/abcbfa}

\bibitem[{{Haurwitz}(1940)}]{1940TrAGU..21..262H}
{Haurwitz}, B. 1940, Transactions, American Geophysical Union, 21, 262,
  \dodoi{10.1029/TR021i002p00262}

\bibitem[{{Hindman} {et~al.}(2020){Hindman}, {Featherstone}, \&
  {Julien}}]{Hindman2020}
{Hindman}, B.~W., {Featherstone}, N.~A., \& {Julien}, K. 2020, \apj, 898, 120,
  \dodoi{10.3847/1538-4357/ab9ec2}

\bibitem[{{Hindman} \& {Jain}(2022)}]{2022ApJ...932...68H}
{Hindman}, B.~W., \& {Jain}, R. 2022, \apj, 932, 68,
  \dodoi{10.3847/1538-4357/ac6d64}

\bibitem[{{Hindman} \& {Jain}(2023)}]{2023ApJ...943..127H}
---. 2023, \apj, 943, 127, \dodoi{10.3847/1538-4357/acaec4}

\bibitem[{{Jain} \& {Hindman}(2023)}]{2023ApJ...958...48J}
{Jain}, R., \& {Hindman}, B.~W. 2023, \apj, 958, 48,
  \dodoi{10.3847/1538-4357/acfc24}

\bibitem[{{Kuo}(1949)}]{Kuo1949}
{Kuo}, H.-L. 1949, Journal of the Atmospheric Sciences, 6, 105,
  \dodoi{10.1175/1520-0469(1949)006<0105:DIOTDN>2.0.CO;2}

\bibitem[{{Lantz}(1992)}]{Lantz1992}
{Lantz}, S.~R. 1992, PhD thesis, Cornell University, New York

\bibitem[{{Lawson} {et~al.}(2015){Lawson}, {Strugarek}, \&
  {Charbonneau}}]{2015ApJ...813...95L}
{Lawson}, N., {Strugarek}, A., \& {Charbonneau}, P. 2015, \apj, 813, 95,
  \dodoi{10.1088/0004-637X/813/2/95}

\bibitem[{{Liang} {et~al.}(2019){Liang}, {Gizon}, {Birch}, \&
  {Duvall}}]{2019A&A...626A...3L}
{Liang}, Z.-C., {Gizon}, L., {Birch}, A.~C., \& {Duvall}, T.~L. 2019, \aap,
  626, A3, \dodoi{10.1051/0004-6361/201834849}

\bibitem[{{Lonner} {et~al.}(2022){Lonner}, {Aggarwal}, \&
  {Aurnou}}]{Lonner2022}
{Lonner}, T.~L., {Aggarwal}, A., \& {Aurnou}, J.~M. 2022, Journal of
  Geophysical Research (Planets), 127, e2022JE007356,
  \dodoi{10.1029/2022JE007356}

\bibitem[{{L{\"o}ptien} {et~al.}(2018){L{\"o}ptien}, {Gizon}, {Birch}, {Schou},
  {Proxauf}, {Duvall}, {Bogart}, \& {Christensen}}]{2018NatAs...2..568L}
{L{\"o}ptien}, B., {Gizon}, L., {Birch}, A.~C., {et~al.} 2018, Nature
  Astronomy, 2, 568, \dodoi{10.1038/s41550-018-0460-x}

\bibitem[{{Mack}(1976)}]{Mack1976}
{Mack}, L.~M. 1976, Journal of Fluid Mechanics, 73, 497,
  \dodoi{10.1017/S002211207600147X}

\bibitem[{{Mason}(1975)}]{Mason1975}
{Mason}, P.~J. 1975, Philosophical Transactions of the Royal Society of London
  Series A, 278, 397, \dodoi{10.1098/rsta.1975.0032}

\bibitem[{{Matilsky} {et~al.}(2023){Matilsky}, {Brummell}, {Hindman}, \&
  {Toomre}}]{Matilsky2023}
{Matilsky}, L.~I., {Brummell}, N.~H., {Hindman}, B.~W., \& {Toomre}, J. 2023,
  arXiv e-prints, arXiv:2311.10202, \dodoi{10.48550/arXiv.2311.10202}

\bibitem[{{Matilsky} {et~al.}(2022){Matilsky}, {Hindman}, {Featherstone},
  {Blume}, \& {Toomre}}]{Matilsky2022}
{Matilsky}, L.~I., {Hindman}, B.~W., {Featherstone}, N.~A., {Blume}, C.~C., \&
  {Toomre}, J. 2022, \apjl, 940, L50, \dodoi{10.3847/2041-8213/ac93ef}

\bibitem[{{Matsui} {et~al.}(2016){Matsui}, {Heien}, {Aubert}, {Aurnou},
  {Avery}, {Brown}, {Buffett}, {Busse}, {Christensen}, {Davies},
  {Featherstone}, {Gastine}, {Glatzmaier}, {Gubbins}, {Guermond}, {Hayashi},
  {Hollerbach}, {Hwang}, {Jackson}, {Jones}, {Jiang}, {Kellogg}, {Kuang},
  {Landeau}, {Marti}, {Olson}, {Ribeiro}, {Sasaki}, {Schaeffer}, {Simitev},
  {Sheyko}, {Silva}, {Stanley}, {Takahashi}, {Takehiro}, {Wicht}, \&
  {Willis}}]{2016GGG....17.1586M}
{Matsui}, H., {Heien}, E., {Aubert}, J., {et~al.} 2016, Geochemistry,
  Geophysics, Geosystems, 17, 1586, \dodoi{10.1002/2015GC006159}

\bibitem[{{Papaloizou} \& {Pringle}(1978)}]{1978MNRAS.182..423P}
{Papaloizou}, J., \& {Pringle}, J.~E. 1978, \mnras, 182, 423,
  \dodoi{10.1093/mnras/182.3.423}

\bibitem[{{Philidet} \& {Gizon}(2023)}]{Philidet2023}
{Philidet}, J., \& {Gizon}, L. 2023, \aap, 673, A124,
  \dodoi{10.1051/0004-6361/202245666}

\bibitem[{{Provost} {et~al.}(1981){Provost}, {Berthomieu}, \&
  {Rocca}}]{1981A&A....94..126P}
{Provost}, J., {Berthomieu}, G., \& {Rocca}, A. 1981, \aap, 94, 126

\bibitem[{{Proxauf} {et~al.}(2020){Proxauf}, {Gizon}, {L{\"o}ptien}, {Schou},
  {Birch}, \& {Bogart}}]{2020A&A...634A..44P}
{Proxauf}, B., {Gizon}, L., {L{\"o}ptien}, B., {et~al.} 2020, \aap, 634, A44,
  \dodoi{10.1051/0004-6361/201937007}

\bibitem[{{Roberts}(1968)}]{1968RSPTA.263...93R}
{Roberts}, P.~H. 1968, Philosophical Transactions of the Royal Society of
  London Series A, 263, 93, \dodoi{10.1098/rsta.1968.0007}

\bibitem[{Rossby(1939)}]{Rossby1939RelationBV}
Rossby, C.-G. 1939, Journal of Marine Research, 2, 38

\bibitem[{{Saio}(1982)}]{1982ApJ...256..717S}
{Saio}, H. 1982, \apj, 256, 717, \dodoi{10.1086/159945}

\bibitem[{{Smeyers} {et~al.}(1981){Smeyers}, {Craeynest}, \&
  {Martens}}]{1981Ap&SS..78..483S}
{Smeyers}, P., {Craeynest}, D., \& {Martens}, L. 1981, \apss, 78, 483,
  \dodoi{10.1007/BF00648954}

\bibitem[{{Smith} {et~al.}(2014){Smith}, {Speer}, \& {Griffiths}}]{Smith2014}
{Smith}, C.~A., {Speer}, K.~G., \& {Griffiths}, R.~W. 2014, Journal of Physical
  Oceanography, 44, 2273, \dodoi{10.1175/JPO-D-13-0255.1}

\bibitem[{{Soderlund} {et~al.}(2012){Soderlund}, {King}, \&
  {Aurnou}}]{2012E&PSL.333....9S}
{Soderlund}, K.~M., {King}, E.~M., \& {Aurnou}, J.~M. 2012, Earth and Planetary
  Science Letters, 333, 9, \dodoi{10.1016/j.epsl.2012.03.038}

\bibitem[{{Triana} {et~al.}(2022){Triana}, {Guerrero}, {Barik}, \&
  {Rekier}}]{2022ApJ...934L...4T}
{Triana}, S.~A., {Guerrero}, G., {Barik}, A., \& {Rekier}, J. 2022, \apjl, 934,
  L4, \dodoi{10.3847/2041-8213/ac7dac}

\bibitem[{{Vallis}(2017)}]{2017aofd.book.....V}
{Vallis}, G.~K. 2017, {Atmospheric and Oceanic Fluid Dynamics: Fundamentals and
  Large-Scale Circulation}, \dodoi{10.1017/9781107588417}

\bibitem[{{Waidele} \& {Zhao}(2023)}]{Waidele2023}
{Waidele}, M., \& {Zhao}, J. 2023, \apjl, 954, L26,
  \dodoi{10.3847/2041-8213/acefd0}

\bibitem[{{Wolff} \& {Blizard}(1986)}]{1986SoPh..105....1W}
{Wolff}, C.~L., \& {Blizard}, J.~B. 1986, \solphys, 105, 1,
  \dodoi{10.1007/BF00156371}

\end{thebibliography}
